%% file: Models_JSV_V44.tex
\documentclass{elsart}

\usepackage{amssymb}
\usepackage{amsmath}
\usepackage{amsfonts}
\usepackage{array}
\usepackage{calc}
\usepackage{dashrule}
\usepackage{dcolumn}
\usepackage[T1]{fontenc}
\usepackage{longtable}
\usepackage{setspace}
\usepackage{verbatim}

\usepackage[english]{babel}
\usepackage{indentfirst}
\usepackage{latexsym}
\usepackage{eucal}
\usepackage{theorem}
\usepackage[cmtip,arrow]{xy}
\usepackage{pb-diagram}
\usepackage{epsfig}
\usepackage{rotate}
\usepackage[it]{subfigure}
\usepackage{color}
\usepackage{textcomp}
\usepackage{lscape}
\usepackage{multirow}
\usepackage[section]{placeins}

\usepackage{relsize}
\usepackage{times}

\newcommand{\Commentaire}[1]{}

\newcommand{\idr}[1]{_{\text{#1}}}
\newcommand{\edr}[1]{^{\text{#1}}}
\newcommand{\Eq}[1]{Eq.~(\ref{eq:#1})}
\newcommand{\Fig}[1]{Fig.~\ref{fig:#1}}

\newcommand{\dd}{\text{d}}

\newcommand{\CR}{\\}

\usepackage{hyperref}
\newcommand{\figureplace}{}
\newcommand{\tableplace}{}

\newcommand{\ED}{\end{document}}

\begin{document}

\begin{frontmatter}

\title{Vibroacoustics of the piano soundboard: {R}educed models, mobility synthesis, and acoustical radiation regime}
\author{Xavier Boutillon}
\ead{boutillon@lms.polytechnique.fr}
\address{Laboratoire de Mécanique des Solides (LMS), École polytechnique, CNRS UMR 7649, \mbox{F-91128 Palaiseau Cedex}, France}

\author{Kerem Ege\corauthref{cor1}}
\ead{kerem.ege@insa-lyon.fr}
\address{Laboratoire Vibrations Acoustique, INSA-Lyon, 25 bis avenue Jean Capelle, \mbox{F-69621 Villeurbanne Cedex}, France}
\corauth[cor1]{corresponding author}

\begin{abstract}
In string musical instruments, the sound is radiated by the soundboard, subject to the strings excitation. This vibration of this rather complex structure is described here with models which need only a small number of parameters. Predictions of the models are compared with results of experiments that have been presented in Ege et al. [Vibroacoustics of the piano soundboard: (Non)linearity and modal properties in the low- and mid- frequency ranges, Journal of Sound and Vibration 332 (5) (2013) 1288-1305]. The apparent modal density of the soundboard of an upright piano in playing condition, as seen from various points of the structure, exhibits two well-separated regimes, below and above a frequency $f\idr{lim}$ that is determined by the wood characteristics and by the distance between ribs. Above $f\idr{lim}$, most modes appear to be localised, presumably due to the irregularity of the spacing and height of the ribs. The low-frequency regime is predicted by a model which consists of coupled sub-structures: the two ribbed areas split by the main bridge and, in most cases, one or two so-called cut-off corners. In order to assess the dynamical properties of each of the subplates (considered here as homogeneous plates), we propose a derivation of the (low-frequency) modal density of an orthotropic homogeneous plate which accounts for the boundary conditions on an arbitrary geometry. Above $f\idr{lim}$, the soundboard, as seen from a given excitation point, is modelled as a set of three structural wave-guides, namely the three inter-rib spacings surrounding the excitation point. Based on these low- and high-frequency models, computations of the point-mobility and of the apparent modal densities seen at several excitation points match published measurements. The dispersion curve of the wave-guide model displays an acoustical radiation scheme which differs significantly from that of a thin homogeneous plate. It appears that piano dimensioning is such that the subsonic regime of acoustical radiation extends over a much wider frequency range than it would be for a homogeneous plate with the same low-frequency vibration. One problem in piano manufacturing is examined in relationship with the possible radiation schemes induced by the models.
\end{abstract}


\end{frontmatter}

\input{IntroV44}

\input{PlateV44}
\input{WaveGuideV44}

\input{MobilityV44}
\input{CoincidenceV44}

\section{Conclusion}
The piano soundboard is a challenging vibro-acoustical object: several more or less independent structures, one of them with a complex ribbing system. In order to gain insight on the vibration regimes that were revealed in previous experimental studies, semi-analytical models have been proposed in this paper. These models consider the different parts of the soundboard as elementary structures: homogeneous plates, structural wave-guides, beam. The models have been inspired by the observation of experimental and numerical modal analyses.

The main part of the soundboard -- a more or less regularly ribbed plate -- has been considered as a homogeneous orthotropic plate. The orthotropy ratio obtained after homogenisation is much smaller than that of spruce. The more difficult problem still consists in describing the coupled dynamics of this homogenised plate and the main bridge (a long beam glued on one side of the soundboard). The solution that has been proposed -- two plates on each side of the bridge which couples them -- yields a modal density of the whole soundboard (including the cut-off corners) in good agreement with previous experimental determinations. In order to derive the modal density of the various plates involved in the model in the low-frequency range, the low-frequency correction due to the boundary condition must be calculated. Two extensions to the existing literature on this particular subject had to be derived (non-special orthotropy) or proposed (arbitrary geometry of the contour). It was also observed that pianos in the same range seem to display similar global properties, namely the rigidity of the isotropic plate equivalent to the whole soundboard at low frequencies.

In the high-frequency regime, the dynamics of the soundboard encounters a marked change due to the ribbing system. Also, the slightly irregular spacing of the ribs is very likely to be the cause of the observed localisation of the modes in the direction orthogonal to the ribs. In this regime, a simple model of three coupled structural wave-guides predicts an apparent (or local) modal density in excellent agreement with the experimental observations.

The point-mobility can be predicted by the models described almost everywhere on the soundboard. For points located at the bridge, these models cannot predict the point-mobility and a previously established model describing plate-beam coupling had to be used. Based on the above observation on dynamical similarity between pianos of similar dimensions, a comparison has been made between the characteristic impedance predicted by the models on a piano that we have measured in detail with the impedance that has been measured in detail on a piano for which only the overall dimensions are known. The features of the characteristic impedance, both at the bridge and far from it, compare very well, not only qualitatively but also quantitatively. 

The vibration models which have been derived can also be used to predict the dispersion curves of the structural waves and thus, the dispersion curves of the corresponding radiated acoustical waves. It appears that the ribbing systems considerably extends the subsonic regime of sound radiation, compared to what it would be on a homogeneous plate equivalent to the soundboard at low-frequencies. It also appears that the ratio of the rib-spacing to the thickness of the main wood panel rules the eventual appearance of the alternation of subsonic and supersonic acoustical radiation regimes. Avoiding an intermediate supersonic radiation regime (which would create a non-regular radiation pattern in the treble range of the instrument) seems to rely on a careful adjustment of geometrical parameters to the wood elastic properties.

\section*{Acknowledgements}
This work has been initiated during the PhD of the second author at the LMS, for which he was sponsored by the French Ministry of Research. We express our gratitude to two anonymous reviewers for the quality of their suggestions. We thank Stephen Paulello for sharing his knowledge of piano making with us on a number of questions. Since 2011, the Laboratoire Vibrations Acoustique is part of the LabEx CeLyA ("Centre Lyonnais d'Acoustique", ANR-10-LABX-60).

\begin{appendix}

\input{CorrectionModalDensityOrthoV44}

\input{Equiv_dynrigid}

\end{appendix}

\bibliographystyle{elsart-num}
\bibliography{SoundboardVibration_Article35}

\end{document}

%% file: IntroV44.tex
\section{Introduction}
\label{sec:Intro}
\begin{figure}[ht!]
\begin{center}
\includegraphics[width=0.56\linewidth]{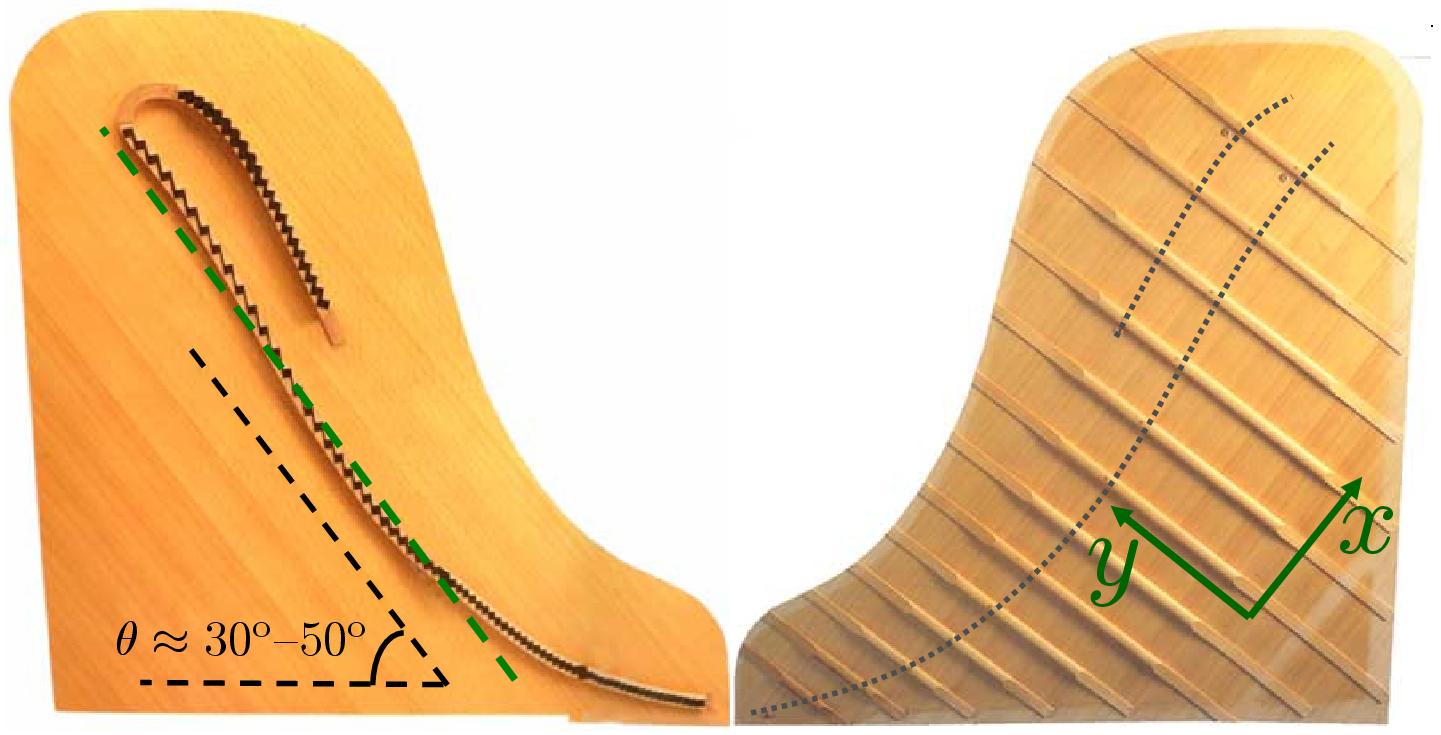}\hspace{0.02\linewidth}%
\includegraphics[width=0.42\linewidth]{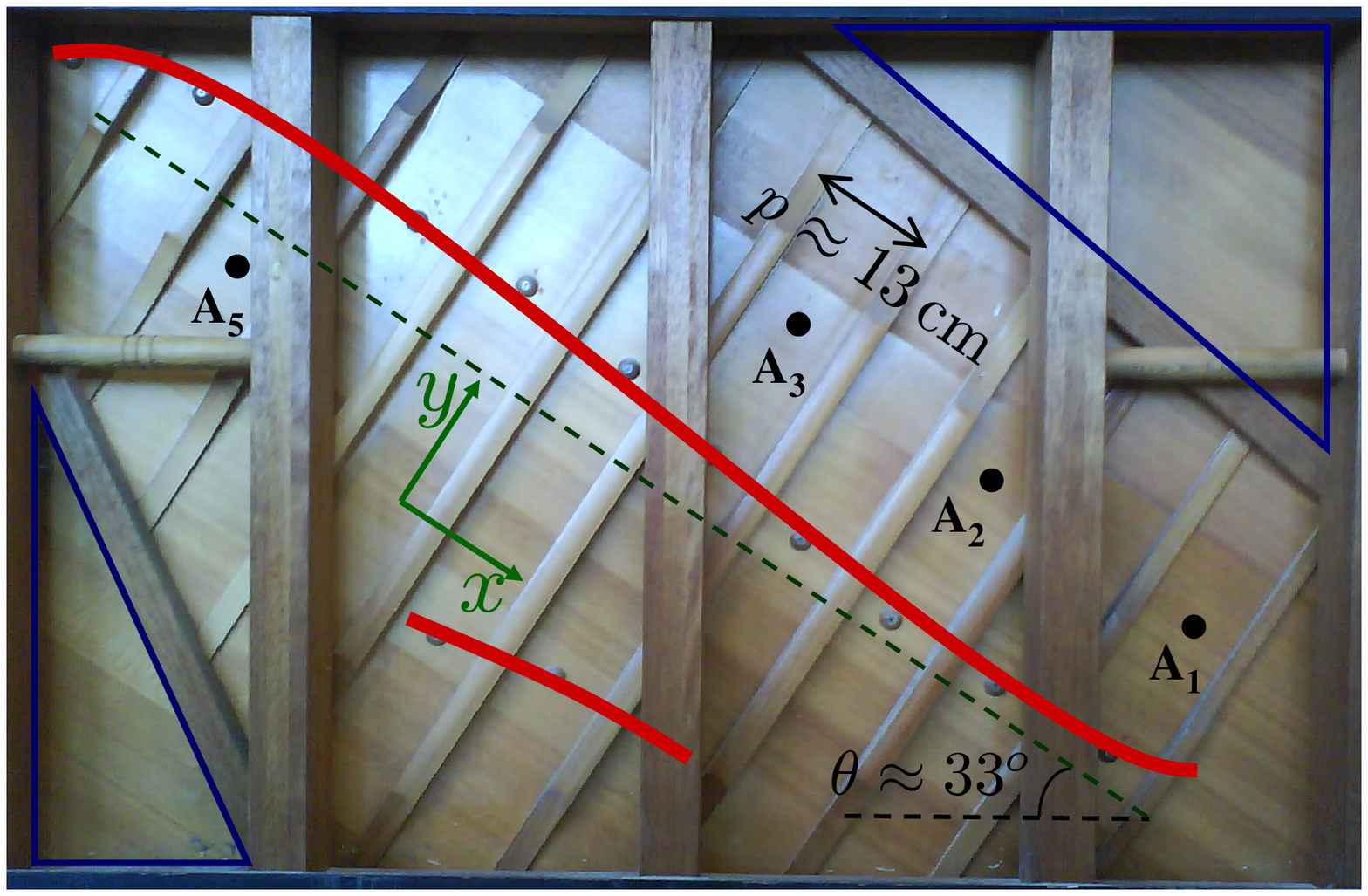}\\
\hspace{.25\linewidth}(a)\hspace{.45\linewidth}(b)
\end{center}
\caption[aaa]{(a): both sides of the soundboard of a grand piano.\CR
(b): the rib side of the soundboard of the Atlas upright piano studied in~\cite{Ege2013a}, with the bridges superimposed as thick red lines and the locations of the accelerometers (in black). The grand soundboard had one cut-off bar, eventually removed. The upright soundboard include one ribbed zone and two cut-off corners (blue-delimited lower-left and upper-right triangles).}
\label{fig:board_threefaces}
\end{figure}

The piano soundboard (\Fig{board_threefaces}) is a large, almost plane, wood-structure. It includes a thin panel made out of glued spruce strips. A series of stiffeners -- the ribs -- are glued across the grain direction of the main panel's wood. The ribs (also made of spruce, sometimes sugar pine) are only roughly equidistant. We define the $x$-direction as the grain direction of the panel and the $y$-direction as that of the ribs. On many pianos, one or two "cut-off" bars (in fir), much wider and thicker than the ribs, form, together with the sides of the soundboard, the so-called "cut-off corners".

Two maple bars -- the bridges -- thicker than the ribs, slightly curved, and eventually partly connected, are glued on the opposite face, roughly in the $x$-direction. The strings of the lower and the upper notes are attached to the (short) bass-bridge and to the main bridge, respectively. The soundboard of upright pianos is rectangular (the strings and the $x$-direction running diagonally). The soundboard of grand pianos looks like a backward slanted "L". The width of the soundboard is more or less 140~cm, corresponding to that of the keyboard. The height or length ranges from more or less 60~cm for very small uprights to more than 2~m for exceptionally large concert grands. The panel thickness is \mbox{$w\approx 8\pm 2$~mm}, the inter-rib distance $p$ ranges from 10 to 18~cm in average (depending on pianos), and is slightly irregular from rib to rib.

\renewcommand{\figureplace}{%
\begin{center}
[\figurename~\thepostfig\ about here (with ref.~\cite{Ege2013a}).]
\end{center}}

\renewcommand{\figureplace}{%
\begin{center}
[\figurename~\thepostfig\ about here.]
\end{center}}

Playing one note corresponds roughly to the following sequence of events: the pianist gives some kinetic energy to the hammer; the hammer escapes its mechanism and interacts very briefly (less than 5~ms) with one, two, or three unison strings; the strings vibrate and exert a localised force at the bridge of the soundboard, which makes the soundboard vibrating and radiating sound toward listeners. In the rest of this paper, "unison strings" will be shortened in "string".

A part of the initial kinetic energy of the hammer is very briefly given to the strings and then slowly transmitted to the soundboard and to the acoustical field. The spectrum of one note (associated with a given pitch) includes a series of almost harmonically related partials: one partial consists of the slightly different modes of the strings and therefore, decays in time with a slow, complex pattern. For a given note, the overall decay-time of each partial must not vary too widely between two consecutive partials. Musically, the timbre must also be balanced from note to note.

The main objective of this paper is to present a semi-analytical model of the soundboard from which one can predict the main characteristics of its vibration when it is excited by one string. More precisely, we focus on the vibration as seen by the string and by the acoustical field. The quantity that represents the coupling between the string and the soundboard is the point-mobility. According to Skudrzyk~\cite{SKU1980}, the average of the real part of the point mobility is directly related to the modal density, which explains in part the emphasis on this parameter throughout the paper. The models presented here have no adjustment parameters and do not rely on the results of dynamical experiments (except for the value of damping). They are meant to explore the changes in vibrational (and partly in radiative) overall properties of the soundboard or in string/soundboard coupling that would be induced by changes in wood characteristics or in the geometry of the various parts of the soundboard. Compared to a finite-element model, our purpose is to provide more understanding and extreme numerical easiness, at the evident price of skipping details, both in space and partly in frequency.

We assume the following approximations:
\begin{itemize}
\item{The soundboard represents a nearly fixed end for the string: strings and soundboard are dynamically weakly coupled and can therefore be modelled independently. In particular, it is considered that they have independent normal modes. Therefore, after the end of the hammer-string interaction, each string vibrates on its normal modes and forces the vibration of the soundboard at frequencies that have no relationship with the soundboard eigenfrequencies.}
\item{Effects due to the shell aspect and to the internal stress of the soundboard will be ignored.}
\item{Only bending waves are considered in the soundboard and in its constituents (plates, bars), with motion in the $z$-direction.}
\item{The mechanical function of the bridge where the string is attached (as seen by the string) is represented by a mechanical admittance, or point mobility:
\begin{equation}
\dot{\xi}(\omega)=Y(\omega)F(\omega)
\end{equation}
where $F$ and $\dot{\xi}$ are, in the Fourier domain, the force exerted by the string and the velocity of the soundboard. For a thin string, $F$ and $\dot{\xi}$ are vectors and $Y$ is a matrix. Only the vector components in the $z$-direction (normal to the plane of the soundboard) and the corresponding matrix coefficient $Y_{zz}$ are considered. The impedance $Z_Q(\omega)$ analysed in Section~\ref{sec:Mobility} must be understood as $Z_Q=Y_{zz}^{-1}$\footnote{By definition, $Y_{zz}$ is $\dot{\xi}$ in response to a unit force $F_z$ combined with zero-forces in the plane $Oxy$ of the soundboard. Therefore, $Z_Q=Y_{zz}^{-1}$ is different, in general, from $Z_{zz}$ which is the response force to a unit imposed $\dot{\xi}$ and zero-velocity (that is: blocked motion) in the $Oxy$ plane. For a detailed discussion of differences between true mobility and true impedance measurements, see~\cite{BOU1999} for example.}.}
\item{All structures (plates, bars) are considered as weakly dissipative, with damping values given by experiments or chosen arbitrarily.}
\end{itemize}

In an experimental study~\cite{Ege2013a}, we analyse several features of the vibration of the soundboard of an upright piano in playing condition: linearity, modal dampings and modal frequencies up to 3~kHz, experimental modal shapes up to 500~Hz, boundary conditions, numerical modal shapes given by a finite-element modelling up to 3~kHz. As far as modal analyses are concerned, two experimental techniques were employed. At low frequencies, the soundboard was hit at 120 points on a rectangular grid covering the whole soundboard and five accelerometers were installed as marked in \Fig{board_threefaces}-b. Results were obtained with a recent modal analysis technique~\cite{EGE2009} based on parametric spectral analysis rather than FFT. Results are good up to about 500~Hz. Above this limit, the energy transmitted by the impact hammer to the structure is limited by either its weight (for light hammers) or by the duration of the impact which is ruled by the first returning impulse from the soundboard, in the order of magnitude of half the longest modal period. Another experimental technique had to be employed, namely to excite the soundboard by an acoustical field. The vibration was measured as before. Although the actual acoustical excitation was continuous in time, it was processed by deconvolution as to make use of the same modal parametric spectral analysis technique as before. However, only the modal frequencies and dampings could be reached by this technique but not the modal shapes. Since the excitation was not local, no point-mobility could be derived with this technique. In these experiments, the measurements are localised responses to the extended excitation by an acoustical field. The piano vibrating scheme of a piano is that of an extended response to a localised excitation. Since these situations are linked by physical reciprocity, results obtained in one situation are the same as in the other one.

\begin{figure}[ht!]
\includegraphics[width=1\linewidth]{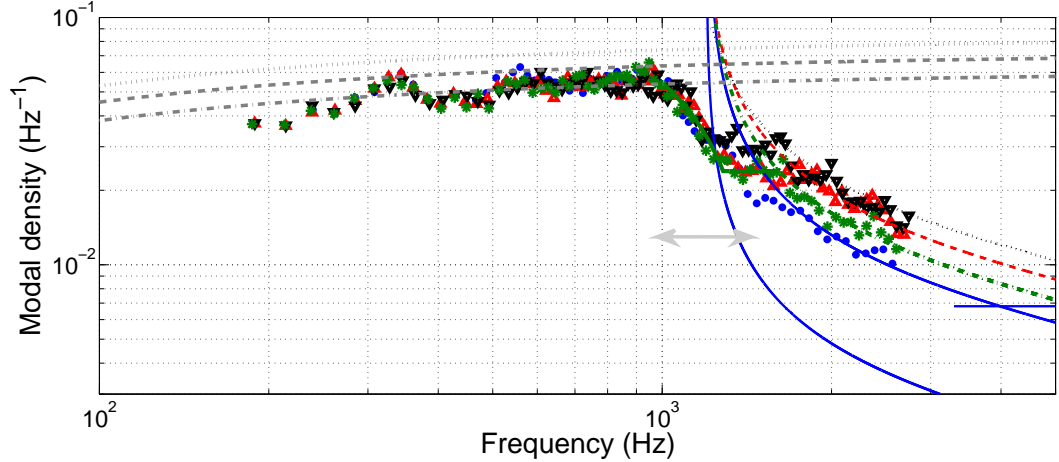}
\caption[aaa]{Modal densities observed on one piano soundboard (dots, data taken from~\cite{Ege2013a}) and evaluated with the model proposed in this article (lines). The estimated values are the reciprocal of the moving average of six successive modal spacings, and reported at the mid-frequency of the whole interval.\CR
Observed values at points \textbf{A}$_\mathbf{1}$ ({\color[rgb]{0,0,1}\tiny{$\bullet$}}), \textbf{A}$_\mathbf{2}$ ({\color[rgb]{1,0,0}$\blacktriangle$}), \textbf{A}$_\mathbf{3}$ ($\blacktriangledown$), and \textbf{A}$_\mathbf{5}$ ({\color[rgb]{0,0.5,0}\scriptsize{$\ast$}}), whose locations are given in \Fig{board_threefaces}~(b). The choice for averaging explains why the first estimated point is well above the first detected mode of the soundboard, at 114~Hz.\CR
Sub-plate model in the low-frequency regime (\S~\ref{sec:LowFreq}): gray lines.\CR
{\color[rgb]{.5,.5,.5}\hdashrule[0.5ex]{5em}{1pt}{2.5mm 1mm .3mm 1mm}}~: "Norway spruce". 
{\color[rgb]{.5,.5,.5}\hdashrule[0.5ex]{5em}{1pt}{2.5mm 1mm}}~: "Sitka spruce".\CR
{\color[rgb]{.5,.5,.5}\hdashrule[0.5ex]{5em}{1.5pt}{.3mm .5mm}}~: "Mediocre spruce" (see Table~\ref{tab:caracmeca_num} for parameter values). Wave-guide(s) model in the high-frequency regime (\S~\ref{sec:HighFreq}), with Norway spruce parameters: colored lines.\CR
Lower {\color[rgb]{0,0,1}\hdashrule[.5ex]{4em}{1pt}{}} line~: modal density of the (1,$n$)-modes in one single inter-rib space enclosing point \textbf{A}$_\mathbf{1}$ (see Eq.~(\ref{eq:ModalDenWG})); the horizontal line at the right hand-side of the figure corresponds to the asymptotic value of the modal density in this wave-guide, with  all possible $(m,n)$-modes.\CR
Group of thin lines: modal density in the set of three adjacent wave-guides.
{\color[rgb]{0,0,1}\hdashrule[0.5ex]{4em}{1pt}{}}~: set of wave-guides in the vicinity of \textbf{A}$_\mathbf{1}$; {\color[rgb]{1,0,0}\hdashrule[0.5ex]{4em}{1pt}{2.5mm 1mm}}~: vicinity of \textbf{A}$_\mathbf{2}$.\CR  {\hdashrule[0.5ex]{4em}{1pt}{0.3mm 0.5mm}}~: vicinity of \textbf{A}$_\mathbf{3}$;\quad {\color[rgb]{0,.5,0}\hdashrule[.5ex]{5em}{1pt}{2.5mm 1mm .3mm 1mm}}: vicinity of \textbf{A}$_\mathbf{5}$.\CR
Thick {\color[rgb]{0,0.5,0}\hdashrule[.5ex]{4em}{1pt}{}}~line: transition between the three-waveguide model and the sub-plate model (\S~\ref{sec:Transition}), for point \textbf{A}$_\mathbf{5}$.
}\label{fig:DensiteModale}
\end{figure}

Many observations can be summarised in \Fig{DensiteModale}, presenting the frequency dependency of the observed modal density, and in the following conclusions:
\begin{enumerate}
\item{The vibration is essentially linear.}
\item{Except at very low frequencies, the boundary conditions are fixed.}
\item{Below \mbox{$\approx$ 1.1 kHz}, the modes extend over the whole soundboard. The modal density slowly increases  and tends towards a constant value of about \mbox{$0.06$ modes Hz$^{-1}$}. The evaluation of the modal density is the same everywhere across the soundboard.}
\item{For frequencies above \mbox{$\approx$ 1.1 kHz}, the ribs confine wave propagation and inter-rib spaces appear as structural wave-guides (wave-number selection in the $x$-direction), as already shown by Berthaut~\cite{BER2004}, \S~V.5. Moreover, modal shapes appear in Fig.~15 of \cite{Ege2013a} as localised in restricted areas of the soundboard, presumably due to the slightly irregular spacing and geometry of the ribs, in all pianos that we have observed. Localisation implies that the number of detected modes per frequency band may vary across the soundboard: at a given place an \emph{apparent} modal density is estimated. This phenomenon is further discussed at the beginning of Section~\ref{sec:HighFreq}.}
\item{The loss factor is $\approx 2\%\,\pm1\%$ over several kHz, without strong systematic variation.}
\end{enumerate}

\renewcommand{\figureplace}{%
\begin{center}
[\figurename~\thepostfig\ about here (with ref.~\cite{Ege2013a}).]
\end{center}}

\renewcommand{\figureplace}{%
\begin{center}
[\figurename~\thepostfig\ about here.]
\end{center}}

In Section~\ref{sec:LowFreq}, devoted to the low-frequency regime, we model the ribbed part of the soundboard as a homogeneous plate and the whole soundboard as a set of sub-plates with clamped boundary conditions, and one bar representing the main bridge. In Section~\ref{sec:HighFreq}, devoted to the high-frequency regime, we model the soundboard as a set of three structural wave-guides and also describe the transition with the sub-plate model. In Section~\ref{sec:Mobility}, we derive the local mobility of the soundboard from the model. In Section~\ref{sec:RadEff}, we analyse the implications of the model on the acoustical radiation and we discuss the consequences in terms of piano manufacturing.

%% file: PlateV44.tex
\section{Low-frequency behaviour: the sub-plate model}
\label{sec:LowFreq}
\subsection{General presentation}
\label{sec:SubplatePresentation}
\begin{figure}[ht!]
\begin{center}
\includegraphics[width=0.5\textwidth]{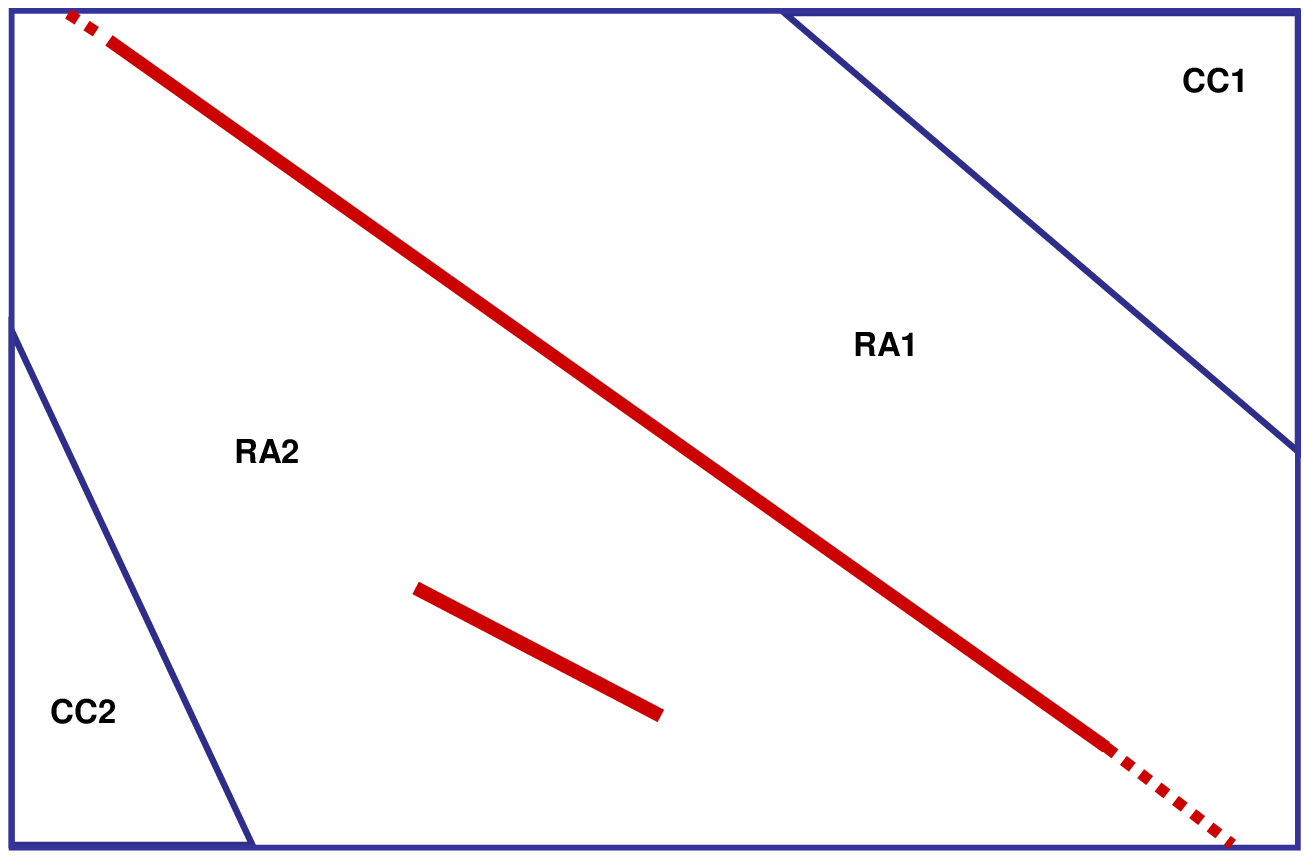}%
\end{center}
\caption[aaa]{Geometry of the subplates: cut-off corners (CC1 and CC2), ribbed areas of the soundboard (RA1 and RA2, the latter including the bass bridge). The main bridge is considered as one of the sub-structures. The orthotropy angle $\theta_\perp$ (defined here as the angle between the keyboard-side of the soundboard and the main axis of orthotropy) is equal to $-32.5^\text{o}$. The panel thickness and cut-off corners thickness are constant and equal to $w$=8mm.}
\label{fig:PlateGeo}
\end{figure}

On almost all pianos, the main bridge extends throughout the entire soundboard and nearly reaches the rims. As shown in \Fig{PlateGeo}, we consider four zones of the soundboard: each side of the main bridge, with a fictitious extension up to the rim (ribbed areas RA1 and RA2) and the two cut-off corners (CC1 and CC2, unribbed except on some large grand pianos). On some pianos, mostly grands, only one or even no cut-off corner may exist. It is assumed that the cut-off bars and the main bridge are stiff and massive enough to keep modes nearly confined within one of these regions. As discussed at the end of this section, this hypothesis is not fulfilled by the bridge, for the first modes. In the model proposed for the low-frequency regime (below $\approx$~1.1~kHz), the main bridge is also considered as a vibrating structure. The bass bridge is considered as adding mass to the ribbed area (being short and thick, its first eigenfrequency is relatively high). The dynamics of the cut-off bars is ignored. The main bridge and the different regions of the soundboard are considered as weakly coupled homogeneous structures. We tested the model by comparing the predicted and the measured modal densities. In the hypothesis of weakly coupled subsystems, the modal density of the soundboard $n(f)$ is the sum of the modal densities of each structure considered separately (\cite{Mac2003}, Eq.~(30), referring to \cite{COU1953}, Chapter~VI, \S1.3):
\begin{equation}
n(f)=n\idr{main bridge}\,+\,n\idr{RA1}\,+\,n\idr{RA2}\,+\,n\idr{CC1}\,+\,n\idr{CC2}
\label{eq:ModalSum}
\end{equation}

Given the hypotheses presented in Section~\ref{sec:Intro}, the plates that compose the piano soundboard are characterised by their surface densities $\mu=\rho h=M/A$ (in generic terms), their rigidities $D=\dfrac{E h^3}{12(1-\nu^2)}$ (\emph{idem}) or dynamical rigidities $\overline{D} = \dfrac{D}{\mu}$, their areas $A$ and their shapes and boundary conditions. As mentioned above, the surface density of RA2 includes the mass of the bass bridge: $\mu\idr{RA2}=(M\idr{RA2}+M\idr{bass bridge})/A\idr{RA2}$.

The cut-off corners are modelled as orthotropic plates (see Table~\ref{tab:caracmeca_num} and \Fig{PlateGeo} for the values of the parameters).

\renewcommand{\tableplace}{\begin{center}[\tablename~\theposttbl\ about here (with refs.~\cite{BER2003,BER2004,HAI1979}).]\end{center}}

\begin{table}[ht!]
\hspace{-0.07\linewidth}
\begin{center}
\begin{tabular}{|c||c|c|c|c|c|c|c|c|c|c|c|}
\hline
\  & $E_\text{L}$ (GPa)& $E_\text{R}$ (GPa)& $G_\text{LR}$ (GPa)& $\nu_\text{LR}$ & $\rho$~(kg~m$^{-3}$)\\ 
\hline

\hline
"Sitka spruce"& 11.5 & 0.47 & 0.5 & 0.3 & 392\\
\hline
"Mediocre wood"& 8.8 & 0.35 & 0.4 & 0.3 & 400\\
\hline
"Norway spruce"& 15.8 & 0.85 & 0.84 & 0.3 & 440\\
\hline
Fir & 8.86 & 0.54 & 1.6 & 0.3 & 691\\
\hline
Maple &10 &2.2 &2.0 &0.3 &660\\
\hline
\end{tabular}
\end{center} 
\medskip
\caption[aaa]{Mechanical characteristics of spruce and fir species selected for piano soundboards. The data of the first and fourth lines are given by Berthaut~\cite{BER2003} with methodology given in ~\cite{BER2004} \S~V.2.1, those of the second line by French piano maker Stephen Paulello, and the others by Haines~\cite{HAI1979}. The subscripts "$_\text{L}$" and "$_\text{R}$" stand for "longitudinal" and "radial" respectively. The radial and longitudinal directions refer to how strips of wood are cut and correspond to the "along the grain" and the "across the grain" directions respectively. $\nu_\text{LR}$ is called the principal Poisson's ratio. \CR In the geometry of the soundboard, the $x$- and $y$- directions correspond to $_\text{L}$ and $_\text{R}$ respectively for the spruce panel: $E_x=E\idr{L},\ E_y=E\idr{R},\ G_{xy}=G\idr{LR}$.}
\label{tab:caracmeca_num}
\end{table}
\renewcommand{\tableplace}{\begin{center}[\tablename~\theposttbl\ about here.]\end{center}}

Each side of the main ribbed zone of the soundboard is considered as a homogeneous orthotropic plate with similar mass, area, and boundary conditions. Homogenisation is done according to Berthaut (Appendix of~\cite{BER2003}), with values of the densities, elastic moduli, and principal Poisson's ratios of the wood species given in Table~\ref{tab:caracmeca_num}. The choice for the values of the elastic constants and densities of the woods is discussed in \S~\ref{sec:LowFreqDiscussion}. Since the ribs are slightly irregularly spaced along the $x$-direction and have varying heights in the $y$-direction, we adopt the approximation  that the flexibilities of the equivalent plate (inverse of rigidities) are the average flexibilities in each direction. In the piano that we have observed, the orthotropy ratio $D_x\edr{H}/D_y\edr{H}$ of the homogenised plate is only \mbox{$\approx$ 1.4}.

The frequency limit of the low-frequency regime is reached when the ribbed area of the soundboard cannot be considered as homogeneous. This occurs when the wavelength in the spruce panel (considered without ribs) becomes comparable to the inter-rib space $p$. Given the generic dispersion equation \Eq{DispersOrtho} in an orthotropic plate, it comes:
\begin{equation}
\label{eq:fgs}
\omega\idr{g}\edr{s}\,=\,\left(\dfrac{\pi}{p}\right)^2\,\overline{D_x\edr{panel}}^{1/2}
\end{equation}
and the frequency limit of the regime \mbox{$f\idr{lim}=\min(f\idr{g}\edr{s})$} is approximately 1.1 kHz.

\subsection{Modal densities of the separate elements}
In an orthotropic plate with thickness $h$, density $\rho$, Young's moduli $E_x$ and $E_y$, orthotropic angle $\theta_\perp\neq0$ (defined as the angle between the long side of the rectangular plate and the main axis of orthotropy, see Fig.~\ref{fig:angle_ortho}), shear modulus $G_{xy}$, principal Poisson's ratio $\nu_{xy}$ and modelled by the Kirchhoff-Love theory, the dispersion equation writes, in polar coordinates:
\begin{equation}
\label{eq:DispersOrtho}
k^4\,\left[D_x\,\cos^4(\theta-\theta_\bot)+2D_{xy}\,\cos^2(\theta-\theta_\bot)\,\sin^2(\theta-\theta_\bot)+D_y\,\sin^4(\theta-\theta_\bot)\right]=\rho\,h\,\omega^2
\end{equation}
with:
\begin{equation}
\label{eq:rigiditiesortho}
\left\{%
\begin{array}{rlrl}
D_x&=\dfrac{E_x h^3}{12(1-\nu_{xy}\nu_{yx})} &D_{xy}&=\dfrac{\nu_{yx}E_x h^3}{12(1-\nu_{xy}\nu_{yx})}+\dfrac{G_{xy}h^3}{6}\\
D_y&=\dfrac{E_y h^3}{12(1-\nu_{xy}\nu_{yx})} &\qquad\nu_{yx}E_x&=\nu_{xy}E_y
\end{array}\right.
\end{equation}

We adopt the following notations\footnote{$\zeta$ is the square root of the orthotropy ratio. $\gamma$ is called the orthotropy parameter. The orthotropy is said elliptic when $\gamma=1$. For most materials\cite{WIL1968}, $\gamma$ is less than 1.}
\begin{align}
\zeta&=\sqrt{\dfrac{E_x}{E_y}}\quad\Longrightarrow\quad \dfrac{D_x}{D_y}=\zeta^2\label{eq:zetaDef}\\
\gamma&=\dfrac{D_{xy}}{\sqrt{D_x D_y}}\label{eq:gammaDef}\\
\alpha^2&=\dfrac{1}{2}\,-\,\dfrac{D_{xy}}{2\sqrt{D_xD_y}}=\dfrac{1}{2}(1-\gamma)\label{eq:AlphaDef}
\end{align}

It is shown in the Appendix~\ref{sec:AppModalOrtho} that the asymptotic modal density 
of a rectangular orthotropic plate is independent of the orthotropy angle $\theta_\bot$:
\begin{align}
n_{\infty,\text{orth}}&=\dfrac{A}{\pi}\,\sqrt{\dfrac{\zeta\,\rho\,h}{D_x}}\,F(\alpha)=\dfrac{A}{2\overline{D_x}^{1/2}}\dfrac{\zeta^{1/2}\,F(\alpha)}{F(0)}=\dfrac{A}{2\overline{D_x}^{1/4}\overline{D_y}^{1/4}}\dfrac{F(\alpha)}{F(0)}\label{eq:ModalDenOrtho}\\
\text{with}\quad F(\alpha)&=\int_0^{\pi/2}{\left(1-\alpha^2\sin^2\theta\right)^{-1/2}\text{d}\theta}
\end{align}

We did not find in the literature an established formula for the low-frequency correction accounting for the boundary conditions of an orthotropic plate with arbitrary $\theta_\bot$. Since it is a problem of practical importance (many modern materials are orthotropic and modal analysis is applicable in the low-frequency range), we give a calculation of this correction in the Appendix~\ref{sec:AppModalOrtho}, for the case of a rectangular plate.

\begin{equation}
n(f)=n_\infty\left(1+\cfrac{\epsilon\tilde{L}}{\sqrt{4\pi A\overline{f}}}\right)
\end{equation}
where $\overline{f}=f\,n_\infty$ and $\tilde{L}$ are given by \Eq{ContourCorrRect}. As for an isotropic plate, the correction is negative for constrained boundary conditions ($\epsilon=-1$). For an arbitrary contour geometry, we propose \Eq{ContourCorrGen} as a generalised expression of  \Eq{ContourCorrRect} for $\tilde{L}$.

The main bridge is modelled by a bar of length $L\idr{b}$ and dynamical rigidity $\overline{D\idr{b}}$, its modal density is independent $n\idr{b}(f)$ of the boundary conditions~\cite{XIE2004}:
\begin{equation}
n\idr{b}(f)\,=\,\dfrac{L\idr{b}}{\overline{D\idr{b}}^{\,1/4}\,(2\pi\,f)^{1/2}}
\end{equation}

\subsection{Discussion}
\label{sec:LowFreqDiscussion}

For a given geometry, the model presented above is predictive only if the density and elastic parameters of wood are known. This was not the case for the piano that we have analysed experimentally. In~\cite{Ege2013a}, we presented results of a finite-element model with various values for wood parameters. Even though the values of elastic parameters, on one hand, density on the other hand, may display large variations (say, up to 40\%), the span of their ratio is much more restricted since, for a given species, denser comes along with stiffer. We present in \Fig{DensiteModale} the modal densities predicted by \Eq{ModalSum} and the models derived above for the three sets of values indicated in Tab.~\ref{tab:caracmeca_num}. The values predicted by the model are systematically higher than those displayed by the FEM, which is to be expected since finite-element models have generally a stiffening bias.

As shown by the modal shapes displayed in Fig.~14 of \cite{Ege2013a}, considering the main bridge as a separation between two zones of the soundboard is not a valid hypothesis for the very first modes but is acceptable as early as 250 Hz. Also, the boundary conditions for the very first modes are not fully constrained. Since assigning a numerical value to the modal density requires averaging, this concept is not applicable to the lowest modes anyway. Upper in frequency, some modes are confined to one side of the bridge whereas others extend on both sides, the bridge representing a nodal line. Therefore, the assumption of separate coupled plates might appear as not fully valid. However, weakly coupled plates or one plate including both yield almost the same asymptotic modal density since $n_\infty(f)$ is proportional to the surface of the plate. Only the low-frequency correction would differ since the overall perimeter is less than the sum of the two perimeters. The alternate model of a plate stiffened by a bar (the bridge) presented in Section~\ref{sec:Mobility}, and others, also presented in~\cite{BOU2012}, do not exhibit better matches with experimental results of the observed modal density in the low-frequency regime.

The influence on the modal density of the internal stress imposed to the soundboard by downbearing by the strings and constrained boundary conditions at the rim has not been modelled here. The FEM modelling presented in~\cite{Ege2013a} showed that the magnitude of these effects is small, in the range of the approximations made in this paper. Including these effects in the model, presumably by an approximate analytical approach, is left for future research.

%% file: WaveGuideV44.tex
\section{High-frequency behaviour: coupled wave-guides}
\label{sec:HighFreq}
For frequencies above \mbox{$\approx$ 1.1 kHz}, mode counting yields different results depending on the way modes are detected. As shown by numerical estimations of modes in Fig.~16 of \cite{Ege2013a}, counting all the modes of the soundboard results in a modal density continuing the low-frequency trend. However, counting only the modes detected at one given point results in a modal density $n(f)$ depending on the point where it is evaluated and decreasing with frequency. It must be noticed in \Fig{DensiteModale} that the apparent abruptness in the fall of $n(f)$ depends on how it has been estimated: if the average mean had been calculated over less than 6 modal spacings (\mbox{$\approx$ 120 Hz}), the curve would have been less regular in general and the abruptness more pronounced. However, the main phenomenon responsible for the dependency of $n(f)$ on the point where it is evaluated is the localisation of modes in the $x-$direction, as shown in~\cite{Ege2013a}. For non-localised modes (low-frequency regime), the vibration has the same order of magnitude everywhere, except in very restricted areas (nodes). For localised modes (high-frequency regime), the amplitude of vibration outside the region of localisation decreases rapidly with distance (localised modes are associated with evanescent waves). Therefore, the detection or the non-detection of a localised mode in a given point is much more robust to measurement conditions (exact position of the measuring device or excitation, signal-to-noise ratio, etc.) than it would be for an non-localised mode\footnote{The non-detection of a non-localised mode requires that the measuring device be set exactly at a node. Additionally, the amplitude at the node depends strongly on damping.}. Altogether, the observations and assumptions presented above for the high-frequency regime seem reasonable enough to refer to $n(f)$ as \emph{the} apparent local modal density.


It is shown below that the confinement of the waves between ribs (wavenumber selection) together with localisation are responsible for the frequency dependency of $n(f)$ above $f\idr{lim}$. The vibration inside one wave-guide is described in \S~\ref{sec:OneWaveguide}, the association of adjacent wave-guides is discussed in \S~\ref{sec:WaveGuideDiscussion} and a transition zone with the low-frequency regime is proposed in Section~\ref{sec:Transition}.

\subsection{The wave-guide model}
\label{sec:OneWaveguide}
\begin{figure}[ht!]
\begin{center}
\includegraphics[width=0.7\linewidth]{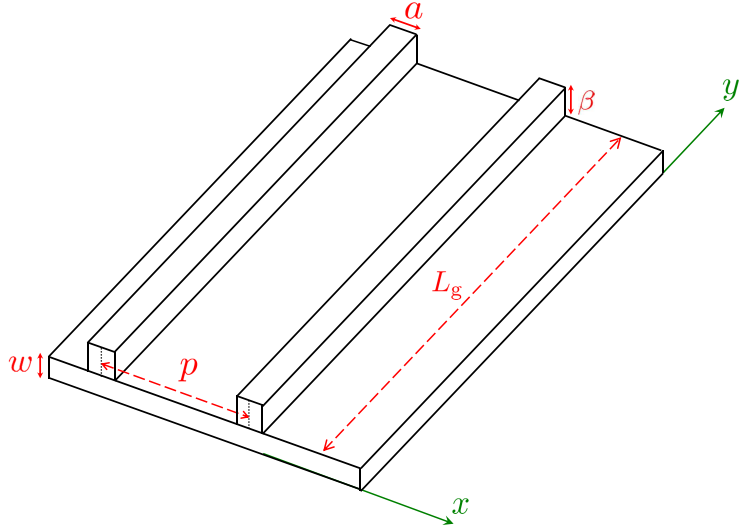}
\end{center}
\caption{Partial scheme of the soundboard between two successive ribs. The thickness of the spruce panel is $w$, the height of the rib is $\beta$, the width of the rib is $a$. $L\idr{g}$ is the average length of one inter-rib region (inter-rib spaces are not rectangular) and varies considerably among inter-rib spaces.}
\label{fig:guidedonde}
\end{figure}

One inter-rib region, schematically represented in \Fig{guidedonde}, behaves like an orthotropic plate of high aspect ratio, with special orthotropy. It is limited in width by the ribs and in length by the rim of the soundboard or by the cut-off bars. As a structure coupled to the rest of the plate, the rib should normally be considered with its full dynamics, including rotation. It is assumed here that, above $f\idr{lim}$, the ribs are heavy enough to impose a nearly fixed condition to the bending transverse waves in the inter-rib region. Although this is true to a lesser extent, it is also assumed that torsion in the ribs do not influence significantly these bending waves. In other words, we assume that the dynamics of the ribs can be ignored and that they represent hinged lines for the vibration in each inter-rib space. The propagation model is that of a structural wave-guide where transverse modes, with discrete wavenumbers $k_{x,m}=m \dfrac{\pi}{p}$, propagate in the $y$-direction.

For a given $k_{x,m}$, the dispersion law (\ref{eq:DispersOrtho}) in the orthotropic portion of the panel between two ribs becomes an equation on $k_y$ only:
\begin{equation}\label{eq:dispersguide}
k_y^4+k_y^2\,\dfrac{2D_{xy}}{D_y}\,k_{x,m}^2+\cfrac{D_x}{D_y}\,k_{x,m}^4-\cfrac{\rho\,h\,\omega^2}{D_y}=0
\end{equation}
where the $D_i$ coefficients are given in Eqs.~(\ref{eq:rigiditiesortho}). With $k_p=\dfrac{\pi}{p}$ and notations introduced in Eqs.~\ref{eq:fgs}, \ref{eq:zetaDef} and \ref{eq:gammaDef}, it comes:
\begin{equation}
k_y^4\,+\,2\zeta\gamma\,k_{x,m}^2\, k_y^2\,+\,\zeta^2\,k_{x,m}^4\left(1-\cfrac{\omega}{m^2\omega\idr{g}\edr{s}}\right)^2=0
\end{equation}

Introducing the $m$-dependent normalisation relationships:
\begin{equation}
\tilde{k}_{y,m}=\dfrac{k_y}{m\,k_p}\qquad\tilde{\omega}_m=\dfrac{\omega}{m^2\,\omega\idr{g}\edr{s}}
\label{eq:Normalisation}
\end{equation}
a dispersion equation of propagating waves in the wave-guide, identical for all transverse modes, is obtained:
\begin{equation}
\tilde{k}_{y,m}^2=\zeta\left[\sqrt{\tilde{\omega}_m^2\,-\,(1-\gamma^2)}\,-\,\gamma\right]\label{eq:DispersWGnorm}
\end{equation}

Each angular frequency $m^2\,\omega\idr{g}\edr{s}$ appears as low cut-off angular frequency associated with the $m$-th transverse mode propagating in the wave-guide of width $p$. The dispersion curves $k_y(f)$ for the two first propagating transverse modes $m=1$ and $m=2$ of an inter-rib space with \mbox{$p=13$ cm} are represented in \Fig{dispers_guide}. They differ noticeably from the succession of pass-bands (separated by stop-bands) that is observed in a more general treatment of the dynamics of the ribbed panel (see~\cite{MAC1980} for example).

In the piano soundboard, the wavenumbers $k_{y,n}$ (in the $y$-direction, parallel to the ribs) are determined by the length $L\idr{g}$ of the wave-guide and by the boundary conditions at the soundboard rim or at the cut-off bars. The ribs defining an inter-rib space have not the same lengths. We assume clamped boundary conditions (as in Section~\ref{sec:LowFreq}) and take for $L\idr{g}$ the length of the longest rib. The wavenumbers $k_{y,n}$ are thus approximated by \mbox{$\left(n+\dfrac{1}{2}\right)\dfrac{\pi}{L\idr{g}}$} with \mbox{$n\in\mathbb{N}^*$}. The modal density in the wave-guide, defined as the reciprocal of the interval between two successive modal frequencies, can be estimated analytically by the usual method, briefly outlined below.

For a given \mbox{$k_{x}=m\,k_p$}, the $(m,y)$-modes with angular frequency less than a given value $\omega^\star$, have the eigen-wavenumbers less than $k_y^\star$, given as a function of $\omega^\star$ by Eqs.~(\ref{eq:Normalisation}) and (\ref{eq:DispersWGnorm}). In the $k_y$-space, these modes occupy the length  $|k_y^\star|$. Since each mode occupies a segment of length $\dfrac{\pi}{L\idr{g}}$, there are \mbox{$N^\star=k_y^\star L\idr{g}/\pi$} such modes. Differentiating with regard to the frequency yields the modal density:
\begin{align}
n_m(f)&=\dfrac{\Delta N}{\Delta f}=\dfrac{\text{d}N^\star}{\text{d}k_y^\star}\,2\pi\cfrac{\text{d}k_y^\star}{\text{d}\omega^\star}=2\,L\idr{g}\dfrac{\dd k_y}{\dd\tilde{k_y}}\,\dfrac{\dd\tilde{\omega}}{\dd\omega}\,\dfrac{\dd\tilde{k_y}}{\dd\tilde{\omega}}\\
&=\dfrac{L\idr{g}\,p}{\pi\overline{D_x}^{\,1/2}}\ \dfrac{\tilde{f_m}}{m\sqrt{\tilde{f_m}^2\,-(1-\gamma^2)}}\,\left(\dfrac{\zeta}{\sqrt{\tilde{f_m}^2\,-(1-\gamma^2)}\,-\gamma}\right)^{1/2}
\label{eq:ModalDenWG}
\end{align}
where $\tilde{f}_m=\dfrac{f}{m^2 f\idr{g}\edr{s}}$.

For the first transverse mode ($m=1$), the theoretical modal density of one of the wave-guides is reported in \Fig{DensiteModale} (lowest solid thin blue line).

As frequency increases, transverse modes corresponding to all possible values of $m$ gradually appear in the wave-guide:
\begin{equation}
n(f)=\dfrac{L\idr{g}\,p}{\pi\overline{D_x}^{\,1/2}}\ \mathop{\sum}_{m=1}^{+\infty} \dfrac{\tilde{f}_m\,H(\tilde{f}_m-1)}{m\sqrt{\tilde{f}_m^2\,-(1-\gamma^2)}}\,\left(\dfrac{\zeta}{\sqrt{\tilde{f}_m^2\,-(1-\gamma^2)}\,-\gamma}\right)^{1/2}
\label{eq:ModalDensityGuide}
\end{equation}
where $H(u)$ is the Heaviside function. However, since the second transverse mode appears above \mbox{$\approx$ 4.4 kHz}, no jump in $n(f)$ appears in \Fig{DensiteModale}. Those will be seen in the next sections, devoted to the synthesis of the mobility and to the acoustical radiation scheme. 

Asymptotically, the modal density of the wave-guide, with all propagating transverse modes, is that of a narrow orthotropic plate of width $p$ and length $L\idr{g}$ (see \Eq{ModalDenOrtho} with $A=p\,L\idr{g}$), represented as an horizontal line at the right of \Fig{DensiteModale}.

\subsection{Discussion}
\label{sec:WaveGuideDiscussion}
\begin{figure}[ht!]
\begin{center}
\includegraphics[width=0.8\linewidth]{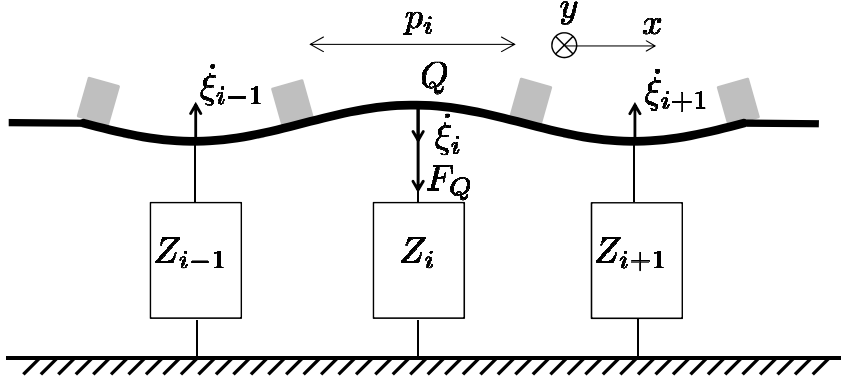}
\caption{Coupling between the bending waves (in the $y$-direction) in wave-guides surrounding the $i$-th one, as excited at $Q$, chosen here at the middle of the $i$-th wave-guide (width $p_i$). Wave-guides are separated by ribs (grey rectangles). When a force $F_Q$ is applied at point $Q$, only the $i-1$-th, the $i$-th, and the $i+1$-th wave-guides are excited (see localisation effect in text) and vibrate in their first transverse mode $k_{x,j}=\dfrac{\pi}{p_j}$. The impedance $Z_j(\omega)$ is the effect of the bending waves dynamics in the $j$-th wave-guide. Coupling results by the summation of the impedances (see text).}
\label{fig:CouplingScheme}
\end{center}
\end{figure}

If the ribs were regularly spaced, the waves (with discrete values of $k_x$) would extend throughout the entire soundboard and the observed modal density would be the same everywhere. As discussed above, irregular spacing is a very probable cause for modal localisation. Inspired by Anderson's theory of (weak) localisation in condensed matter systems, the localisation of vibration in irregular mechanical structures has been extensively studied: see~\cite{HOD1982} for an introduction. We did not find established theoretical means for predicting the localisation areas in the piano soundboard but there is no theoretical reason either for restricting the vibration to one structural wave-guide. Moreover, the shapes of the localised modes reported in Fig.~15 of~\cite{Ege2013a} show that they extend over more than one, but only a very few inter-rib spaces. Following a remark made earlier on the fast spatial attenuation of localised modes, we propose a simplified model in which the vibration extends over three adjacent wave-guides, as represented in \Fig{CouplingScheme}. This assumption is also consistent with the fact that bending waves in adjacent wave-guides are coupled by the finite impedance of bending waves in ribs, particularly near the rim where the ribs are lower (smaller $\beta$).

Altogether, we consider that (a) modes are mainly located in one wave-guide and selected according to the dimensions of this each wave-guide, (b) they have a contribution in the two adjacent wave-guides, (c) more remote regions can be considered as quasi-nodal. When the whole soundboard vibrates under an acoustical excitation, one accelerometer must be sensitive, in this model, to the modes located in three wave-guides. The modal density observed at that point is the sum of the modal densities in each of the three wave-guides, as given by \Eq{ModalDensityGuide}. As shown by \Fig{DensiteModale}, this simplification is in excellent accordance with the experimental results above \mbox{1.5 kHz}. One notices also that the model and the observations coincide on the differences between points. Moreover, multiplying the modal density in one wave-guide by three (not represented in \Fig{DensiteModale}) would not give as a good fit with observations as that obtained here by accounting for the fact that the two neighbouring wave-guides have different lengths.
\vspace{1.45cm}

We analyse now the implications of the three-wave-guide model in terms of the mobility or impedance measured at one point, within the hypotheses and approximations given in the Introduction. Mechanically, coupling between bending waves in the wave-guides operates via the transverse mode. By definition, an impedance $Z_j$ (considered here at the mid-line of wave-guide $j$) is created by the dynamics of the bending waves in the $j$-wave-guide (\Fig{CouplingScheme}, see Section \ref{sec:Mobility} for the analytical treatment). As a thought-experiment, cancelling the Young's modulus $E_y$ (but not $E_x$) and the density $\rho$ in the central and right wave-guides would annihilate forces corresponding to these waves, thus cancelling $Z_i$ and $Z_{i+1}$. In such a circumstance, imposing a motion $\xi_{i,Q}$ at $Q$ would still create the shape of the first transverse mode in the central wave-guide and thus create a motion $\xi_{i-1}$, by coupling between the transverse modes at $y=y_Q$; bending waves would therefore be generated in the left wave-guide. The force needed to establish the transverse mode in the central wave-guide is purely static and the corresponding impedance $Z_{i,E_y=0}$ can be neglected above $\omega\idr{g}\edr{s}$. Assuming perfect coupling and $Q$ at the centre of the central wave-guide, the impedance $Z_Q$ would then be $Z_{i-1}$. In turn, these waves in wave-guide $i-1$ would induce a motion in the central wave-guide, opposite to the motion in the left wave-guide. In normal circumstances (finite $E_y$ and $\rho$), it follows that the 
impedance created at $Q$ is
\begin{equation}
\label{eq:Coupling}
Z_Q=\sum\limits_{j=i-1}^{i+1}Z_j
\end{equation}

The sort of independence between the dynamics in the $x$-direction (dealing with the transverse mode $k_x=\dfrac{\pi}{p}$ and its extension in the two adjacent wave-guides only) and in the $y$-direction (propagation of bending waves in one wave-guide) explains the somewhat unusual circumstance in which modal motions and local forces add, yielding the 
addition of the modal mobilities inside a wave-guide (see Section~\ref{sec:Mobility}) and the addition of wave-guide impedances.

Apparently, the three-waveguide model cannot account for the observed situation of sympathetic strings: playing a low note excites strings of a higher note (with its damper up) having common partials with the low note (usually, the fundamental of the high note). However, this situation implies a number of phenomena that are not examined in this paper. In the sympathetic strings situation, the vibration of the soundboard is forced at frequencies defined by the string(s), therefore combining several modal shapes. In particular, a low-frequency mode (that is: with a low value of the eigenfrequency) does respond, although weakly, to a high frequency excitation and can therefore excite a remote string. Another important feature is probably that modal dampings are much lower for string modes than for soundboard modes. Finally, it must be noticed that the coupling of remote strings occurs via the motion of the main bridge. In the model, its transverse motion is considered as small compared to that of the rest of the soundboard but some significant torsion may well occur. In turn, parametric excitation of transverse string waves by the slight variations in string tension induced by the motion of the bridge in the string direction may also play a role in the generation of the sympathetic strings effect.

\subsection{Transition between the plate and the wave-guide models}
\label{sec:Transition}
Between 1 and 1.5~kHz, a more elaborate model taking into account the dynamics of the ribs would be necessary in order to describe the transition between the sub-plate and the 3-wave-guides models. We present here a cruder description. As explained at the end of \S~\ref{sec:SubplatePresentation}, the sub-plate model breaks at $f\idr{lim}$ (the smallest of $f\idr{g}\edr{s}$), given by \Eq{fgs}), defining the same upper bound of the sub-plate model for all measured points. We define the centre of the transition zone (noted by a double grey arrow in \Fig{DensiteModale}) as the frequency for which the sub-plate model and the three-waveguide model have the same modal density (\emph{e.g.}, at \mbox{$\approx 1310$~Hz} for point \textbf{A}$_\mathbf{5}$, as shown by the intersection of the corresponding modal density curves in \Fig{DensiteModale}). This supposed "centre" also defines, somewhat arbitrarily, the upper bound of the transition zone.

Starting at the upper bound of the transition zone (where the wave-guide model becomes fully valid) and decreasing frequency, the modes are expected to extend gradually throughout the soundboard. Guided by the experimental results, we can expect that this makes the region of the central wave-guide more remote, and therefore, more nodal (for modes located in remote wave-guides) than when modes are localised in the adjacent wave-guides. In consequence, the modal density should be less than what it would be if the model was still holding. Arbitrarily, we consider that the modal density keeps the same value as the frequency decreases, down to the frequency at which the modal density of the single central wave-guide is reached. From there, we took a linear transition toward the lower bound of the transition zone (upper bound of the sub-plate model), at $f\idr{g}\edr{s}$. The transition is represented by the thick plain broken line in \Fig{DensiteModale}.

%% file: MobilityV44.tex
\section{Synthesised mechanical mobility and comparison with published measurements}
\label{sec:Mobility}
In this section, we evaluate the point mobility at different points of the soundboard. The point mobility at the bridge, where a string is attached, describes the coupling between the string and the soundboard. It is a key point for numerical sound synthesis based on physical models and more generally, for the understanding of sound characteristics: crucial musical parameters such as the damping of coupled string modes depend on the mechanical mobility and on the mistuning between unison strings ~\cite{WEI1977}. In very generic terms, the modal density involves a ratio between mass and stiffness whereas mobilities involve their product. A good model should therefore be able to predict both.

Under the assumptions presented in Section~\ref{sec:Intro}, the mobility $Y_Q$ at a point $Q(x_Q,y_Q)$ of the soundboard can be expressed as the sum of the mobility of the normal modes, considered as generalised coordinates, each having the dynamics of a linear damped oscillator\footnote{The convention for time-dependency is $\exp(j\omega t)$.}:
\begin{equation}
\label{eq:drive_admitt_amort}
Y_Q(\omega)=\dfrac{\dot{\xi}_Q(\omega)}{F_Q(\omega)}=j\omega\,\mathop{\sum}_{\nu=1}^{+\infty}\,\dfrac{\overline{\xi_{\nu,Q}}^2}{m_\nu\,(\omega_\nu^2+j\eta_\nu\omega_\nu\omega-\omega^2)}
\end{equation}
where $m_\nu$ is the modal mass, $\eta_\nu$ the modal loss factor, $\omega_\nu$ the modal angular frequency, $\xi_Q$ the transverse displacement at $Q$, and $\overline{\xi}_{\nu,Q}$ the normalised displacement at $Q$ for the mode $\nu$ (modal shape). In this article, the modal mass is defined as the mass that gives the same maximum kinetic energy to the harmonic modal oscillator as that of the actual plate when both vibrate at the corresponding modal frequency with a unit maximum displacement.

With a modal density of roughly \mbox{0.02 Hz$^{-1}$}, about 200 modes are potentially involved in the \mbox{[0,10] kHz} frequency range. The modal description does not bear any musical significance in itself and it is hard to think of cues that would help to sort out this huge number of modal parameters or to establish a hierarchy between them. If the quality of a piano was depending on the particular geometry of (some) modal shapes or on the particular values of (some) modal frequencies, those would be adjusted by piano makers. This is not the case, except perhaps at very particular locations like string-crossing, bridge ends or where the number of unison strings changes. Therefore, it seems reasonable to derive the values of the modal parameters from physical models, which depend on a much smaller number of parameters. We present a mode-by-mode synthesis (\S~\ref{sec:SymthMob}) and a mean-value approach (\S~\ref{sec:AvMob}) based on Skudrzyk's theory~\cite{SKU1980}. 


We are not interested here in specific point-locations. Modal shapes can thus be described by random distributions:
\begin{equation}
\left\{%
\begin{array}{rlrl}
\text{Plate:}\qquad \overline{\xi}_{\nu,Q}&=\sin{2\pi\alpha}\,\sin{2\pi\beta}\\
\text{Wave-guide:}\qquad \overline{\xi}_{\nu,Q}&=\sin(2\pi k_{x,m}\, x_Q)\,\sin{2\pi\beta}
\end{array}\right.
\label{eq:ModalShapes}
\end{equation}
where the random quantities $\alpha$ and $\beta$ are uniformly distributed in [0,1]. For $Q$ near the centre of a waveguide, $\sin(2\pi k_{x,m}\, x_Q)$ is approximately 1 for $m$ odd and approximately 0 for $m$ even. With the chosen definition of the modal mass and normalisation of modal shapes, it follows that all modal masses, whether for a plate or a wave-guide, are given by $m_\nu=M\idr{plate,waveguide}/4$.

\subsection{Synthesised mobility at the bridge}
\label{sec:SymthMob}
In the model described in Section~3, the main bridge is considered as a nodal line for most modes of the rest of the soundboard. Therefore, no sensible value of $\xi_{\nu,Q}$ can be assigned for these modes when $Q$ is on the bridge. We follow an other approximate model given by Skudrzyk~\cite[p.~1129, second approximation presented]{SKU1980} for a plate stiffened by a bar:
\begin{itemize}
\item{The impedance of the soundboard at the bridge is the sum of the impedance of the bridge, considered as a beam (see next points below for its characteristics), and that of the central zone of the soundboard, considered as a plate (idem).}
\item{The coupling between the bridge and the plate is described by a modified modal density of the bridge and by an added stiffness on the plate.}
\item{The modified modal density of the bridge is obtained by adding to the bridge the mass of the  plate.}
\item{In its direction, the bridge imposes its stiffness to the plate (modified orthotropy compared to \S~\ref{sec:SubplatePresentation}, taken arbitrarily as elliptic).}
\end{itemize}

\begin{figure}[ht!]
\centering
\includegraphics[width=.7\linewidth]{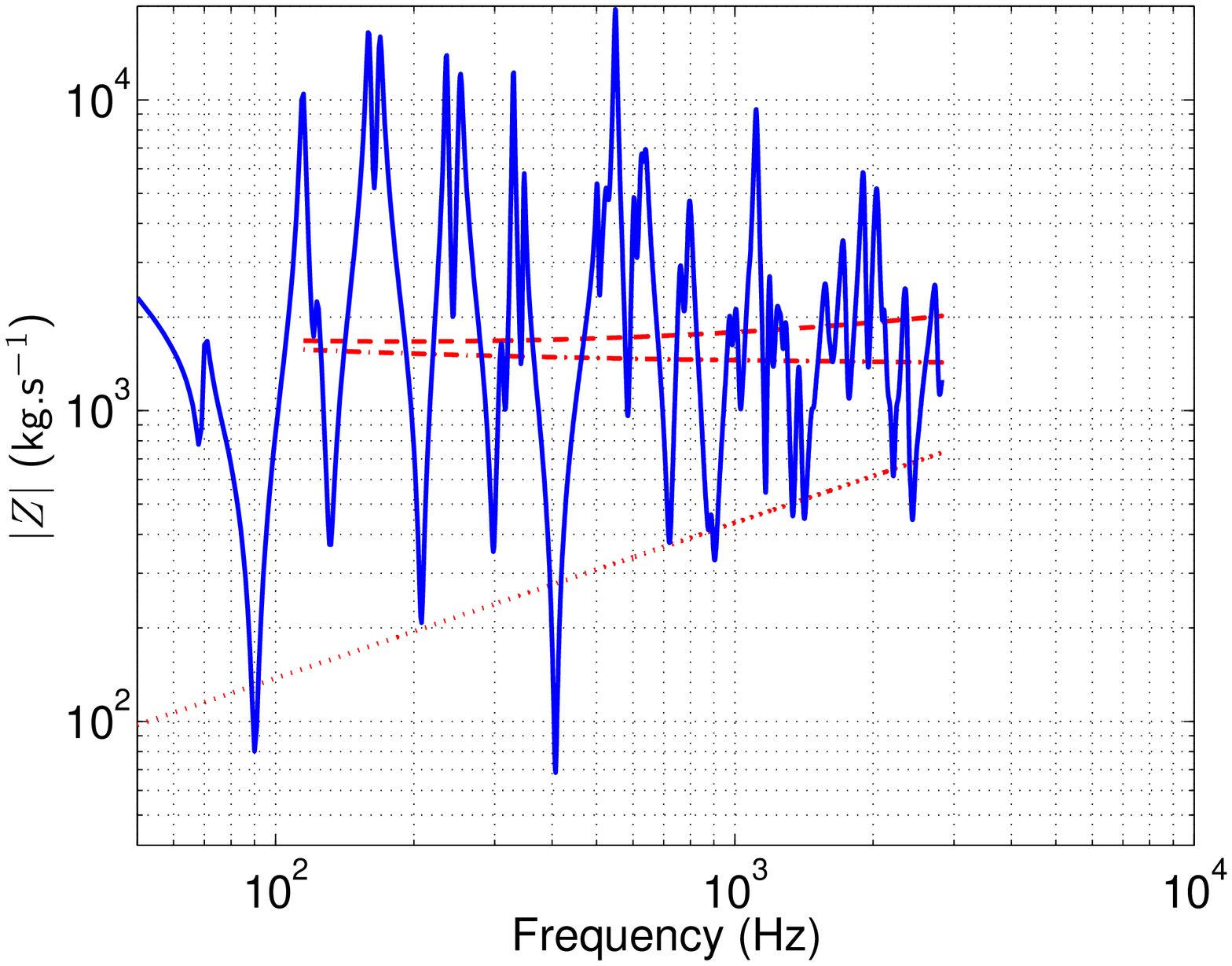}
\caption[bbb]{Synthesised impedance at the bridge (blue solid line) and characteristic impedances (red lines, see \S~\ref{sec:AvMob}).\CR
Dashed line: characteristic impedance of the soundboard $Z\idr{c}=1/Y\idr{c}$, resulting from the sum of the characteristic impedances of the ribbed zone and of the bridge.\CR
Dash-dotted line: characteristic impedance of the ribbed zone of the soundboard, with a stiffness in the bridge direction imposed by the bridge (see text).\CR
Dotted line: characteristic impedance of the bridge considered as a beam mass-loaded by the plate (see text).
}
\label{fig:AdmitBridge}
\end{figure}

A synthesised impedance (reciprocal of the point-mobility) at the bridge is presented in \Fig{AdmitBridge}, according to Eqs.~(\ref{eq:drive_admitt_amort}) and (\ref{eq:ModalShapes}). 

The first two values of the modal frequencies are those measured experimentally (114 and 134 Hz) but any physically reasonable choice would be acceptable since we are not interested in specific modal values. Beyond homogenisation that has been done in the low-frequency regime, a significant degree of irregularity remains in the soundboard, which must be considered here as uncertainty. According to~\cite{WEA1989}, the spacing of the modal frequencies obeys a Rayleigh distribution instead of the Poisson distribution which rules modal spacings of regular structures. The values of the modal frequencies are determined by the random distribution and the modal density given by the model presented in Section 3:
\begin{equation}
\label{eq:SynthModFreq}
f_{\nu+1}\,=\,f_{\nu}+\dfrac{1}{n(f_\nu)}\,r
\end{equation}
where $r$ is a random number with the following probability distrubtion:
\begin{equation}
p\idr{d}(x)\,=\,\dfrac{\pi\,x}{2}\,\exp\left(-\,\dfrac{\pi\,x^2}{4}\right)
\end{equation}
of mean 1 and variance $\dfrac{4}{\pi}-1$. The same randomisation of the modal spacing was chosen by Woodhouse~\cite{WOO2004b} for his statistical guitar\footnote{See Eq.~5, where it seems that $(2/\pi)$ has been written instead of $(2/\pi)^{1/2}$.}. When $Q$ is outside the cut-off corners, it must be considered as a node for the modes of the cut-off corners: the modal density in \Eq{SynthModFreq}) is restricted to that of the ribbed zone of the soundboard. The modal frequencies of the bridge have not been randomised.

In accordance with the experimental values that have been found for the loss factor~\cite{Ege2013a}, we attribute the following values to the modal dampings:
\begin{equation}
\eta_\nu\,=\,\dfrac{\alpha_\nu}{\pi\,f_\nu}\,=\,\dfrac{2.3}{100}r
\end{equation}
where $r$ i a random number following a chi-square probability distribution, as proposed by Burkhardt and Weaver~\cite[Eq. 8]{BUR1996a}. 

As long as the wavelength in the bridge is large compared to the inter-rib spacing, the bridge is coupled to the whole soundboard and the effect of localisation is expected to be lost. It is expected to reappear when the half-wavelength in the bridge becomes equal to the inter-rib spacing. We have limited the frequency range of the synthesis to this limit.

\subsection{Comparison with experiments: the mean-value approach}
\label{sec:AvMob}
\label{sec:Mob_meas}
The only reliable and well-documented point-mobility measurements of a piano soundboard available in the literature are those by Giordano~\cite{GIO1998}. This author presents his measurements in the form of impedances. Giordano's soundboard differs in size from ours by only a few centimetres. We form the hypothesis that these pianos are dynamically comparable. This hypothesis is grounded by the observation that different pianos that have been measured in the literature seem to display comparable equivalent isotropic rigidities (see Appendix~\ref{sec:equiv_dynrigid}). A one-by-one comparison between modes, peaks, etc. between two pianos would be meaningless. We have adopted the mean-value approach of Skudrzyk~\cite{SKU1980}. His theory predicts the value of the geometrical mean of the real part of the mobility of a weakly dissipative structure as a function of the modal density and the mass of the structure. It also shows that weak damping has no influence on the average level of the mobility. An outline and the main results are given below.

For a given mode, the geometrical mean of $\Re(Y_Q(\omega))$ is:
\begin{equation}
\label{eq:GcMode}
G_\nu = \dfrac{n(f)}{4\,M_\nu}\qquad\text{with}\quad M_\nu\,=\,M\,\dfrac{<\xi_\nu^2>}{\xi_{\nu,Q}^2}
\end{equation}
where $M$ is the total mass of the structure.

For a plate or a beam, averaging on the modal shape and then on the modes yields the real part of the so-called characteristic mobility: $$G\idr{c,plate,beam} = \dfrac{n}{4\,M\idr{plate,beam}}$$. The geometrical mean of the imaginary part of the mobility differs between plates and beams. Finally, the characteristic mobilities are:
\begin{equation}
\label{eq:YcP}
Y\idr{c,plate}(f) = \dfrac{n\idr{plate}(f)}{4\,M\idr{plate}} \qquad Y\idr{c,beam}(f)\,=\, \dfrac{n\idr{beam}(f)}{4\,M\idr{beam}}(1\,-\,j)
\end{equation}

In a waveguide, averaging on the modal shapes yields a different result:\\ \mbox{$G\idr{c,guide,m} =\,\epsilon_{Q,m}\dfrac{n\idr{guide}}{2\,M\idr{g}}$}, where, as shown above, $\epsilon_{Q,m}$ is shared by all modes corresponding to a given propagating transverse mode and depends on the location of $Q$. Near the middle of the wave-guide, $\epsilon_{Q,m}$ is approximately 1 for odd values of $m$ and approximately 0 for even values of $m$. All the dispersion branches in the wave-guide (corresponding to the successive $m$-th propagating transverse modes) behave asymptotically like the dispersion equation of a beam (see \Eq{DispersWGnorm} and \Fig{dispers_guide}). We consider therefore that \mbox{$B\idr{c,guide}\approx -G\idr{c,guide}$}. Accounting for the coupling of three wave-guides described by \Eq{Coupling}, it comes:
 \begin{equation}
\dfrac{1}{Y\idr{c,3guides}(f)}\,=\, \mathop{\sum}_{k=1}^{3}\,\left[\mathop{\sum}_{m=1,2,...}\,\dfrac{\,n_{\text{guide,}k}(f)}{2\,M_{\text{guide,}k}}(1\,-\,j)\epsilon_{Q,m}\right]^{-1}
\label{eq:YcW}
\end{equation}

\begin{figure}[ht!]
\centering
\includegraphics[width=.48\linewidth, height=55mm]{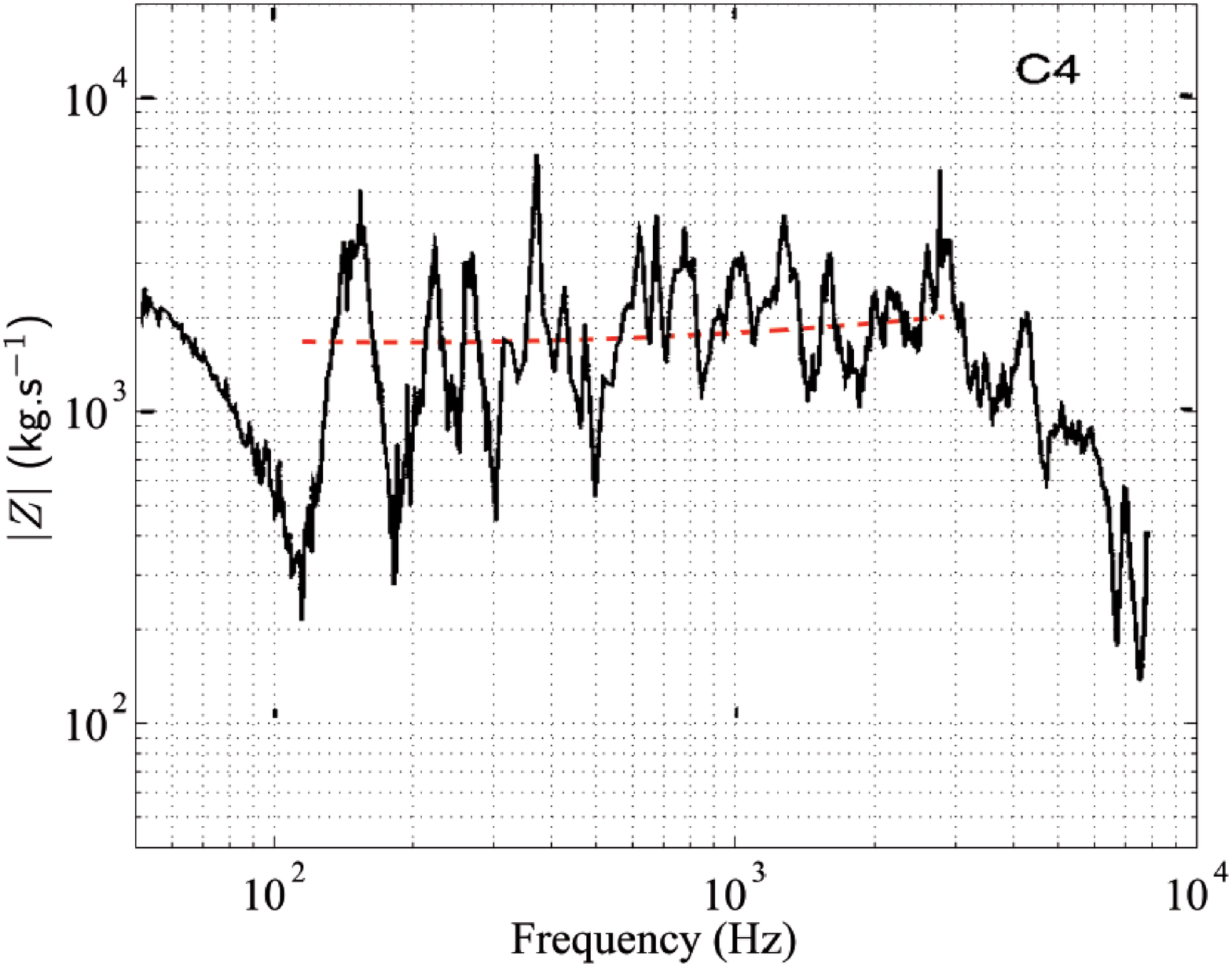}\hspace{.02\linewidth}%
\includegraphics[width=.48\linewidth, height=55mm]{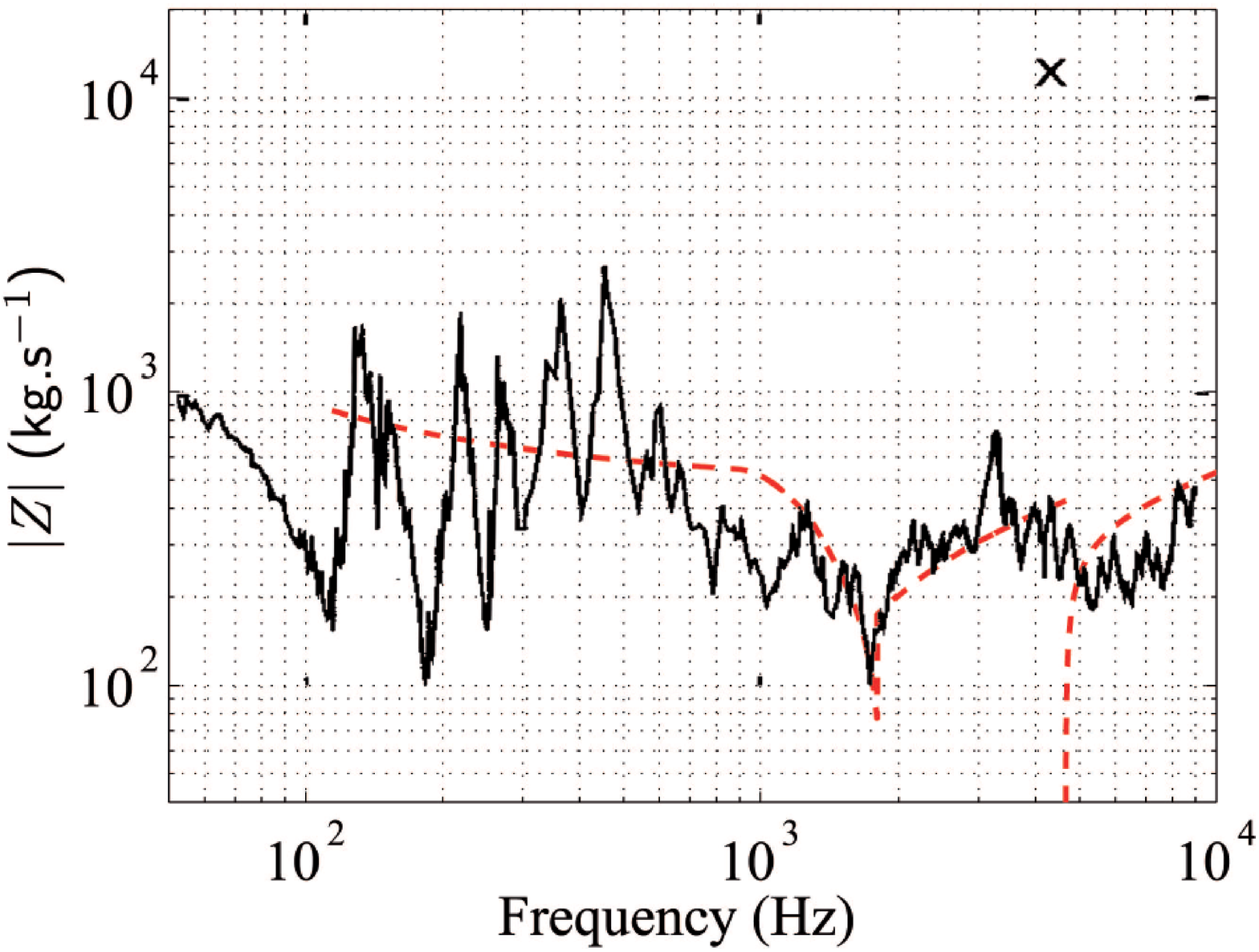}
\caption[bbb]{Magnitude of measured point-impedances by Giordano\cite{GIO1998} (solid black lines) and characteristic impedances modeled on a similar piano (dashed red lines, see text). Left frame: at the bridge, where strings C4 are attached. Right frame: far from the bridge, between two ribs ("X"point in Fig.~1~(a) of~\cite{GIO1998}).
}
\label{fig:CompImpGior}
\end{figure}

At the bridge, the characteristic impedances (reciprocal of the mobilities given by \Eq{YcP}) add, for the beam and plate modified as described in \S~\ref{sec:SymthMob}. They are represented in \Fig{AdmitBridge} and their sum is reported in the left frame of \Fig{CompImpGior}, taken from Giordano.

Far from the bridge (right frame of \Fig{CompImpGior}), the characteristic mobility was computed according to \Eq{YcP} for the low-frequency regime and to \Eq{YcW} for the high-frequency regime, at a point similar to point "X" in Giordano's piano (see Fig.~1~(a) in~\cite{GIO1998}). The low-to-high transition for the modal density is described in \S~\ref{sec:Transition}. At the high-frequency end, the contribution of the second propagating transverse mode is somewhat arbitrary since the precise location along the $x$-axis is unknown. 

Given the approximations made in the models and their application to a piano that we did not measure directly, one may consider that the match is striking, except in the transition zone, as could be expected. The excellent agreement in the upper frequency range may be considered as a partial confirmation of the coupling scheme devised in Section~\ref{sec:WaveGuideDiscussion}. Above the frequency for which half of the wavelength in the bridge becomes comparable to the inter-rib spacing (\mbox{$\approx$3~kHz}), the bridge begins to "see" the ribs. This might be an explanation for the impedance decrease at the bridge, above 3~kHz (left frame of \Fig{CompImpGior}). Again, irregular spacing is likely to cause localisation, coming along with a decrease in impedance (right frame of \Fig{CompImpGior}, above \mbox{$\approx$1~kHz}.

\renewcommand{\figureplace}{%
\begin{center}
[\figurename~\thepostfig\ about here (with ref.~\cite{GIO1998}).]
\end{center}}

\renewcommand{\figureplace}{%
\begin{center}
[\figurename~\thepostfig\ about here.]
\end{center}}

%% file: CoincidenceV44.tex
\section{Some features of the acoustical radiation}
\label{sec:RadEff}
In this section, we examine how the soundboard acoustical radiation differs from the standard radiation scheme of a plain plate with the aim of examining some allegedly critical factors in piano making. For the sake of simplicity, we restrict our attention to the dispersion equations. Acoustical and structural waveumbers are denoted with "a" and "s" superscripts respectively.

\subsection{Radiation regimes}
\label{sec:Coincidence}
\begin{figure}[ht!]
\begin{center}
\includegraphics[width=\linewidth]{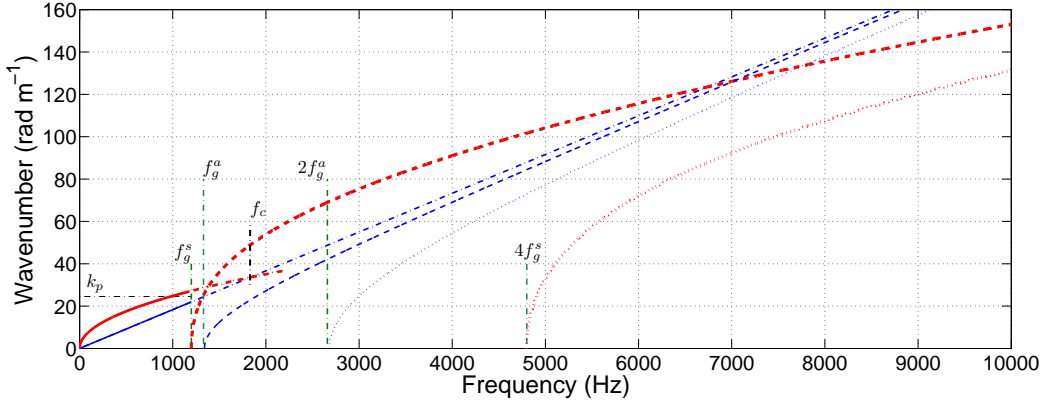}
\caption[asdf]{Dispersion curves of structural and acoustical waves generated inside and radiating outside the piano soundboard: supersonic \emph{vs.} subsonic structural waves. Thick red curves: bending waves in the homogenised plate equivalent to the ribbed zone of the soundboard and in a structural wave-guides. Thin blue curves: corresponding radiated acoustical waves. The acoustical radiation is efficient (supersonic waves) for frequencies at which a blue curve is above the red curve with the same motive.\CR
{\color[rgb]{1,0,0}\hdashrule[.5ex]{4em}{3pt}{}}\ \ and {\color[rgb]{1,0,0}\hdashrule[0.5ex]{5em}{3pt}{2mm 2pt .7mm 2pt}}~: $|k_x\edr{s}(f)|$ for the fastest bending waves ($x$-direction, lowest $f\idr{c}$) in the homogenised plate equivalent to the ribbed zone of the soundboard, respectively below and above $f\idr{g}\edr{s}$ (crossing the dispersion line of plane waves in air at $f\idr{c}$).\CR
{\color[rgb]{1,0,0}\hdashrule[0.5ex]{5em}{3pt}{2.5mm 1mm}}~: $|k_y\edr{s}(f)|$ for bending waves in the wave-guide between the second and third ribs for the first transverse mode of the guide, starting at $f\idr{g}\edr{s}$ (see Eqs.~\ref{eq:fgs} and \ref{eq:DispersWGnorm}).
{\color[rgb]{1,0,0}\hdashrule[0.5ex]{5em}{3pt}{1mm 3mm}}~: $|k_y\edr{s}(f)|$ for bending waves in the wave-guide between the second and third ribs for the second transverse mode, starting at $4\,f\idr{g}\edr{s}$.\CR
{\color[rgb]{0,0,1}\hdashrule[.5ex]{4em}{1pt}{}}\ \ and {\color[rgb]{0,0,1}\hdashrule[0.5ex]{5em}{1pt}{2mm 2pt .7mm 2pt}}~: $|k\edr{a}(f)|$ for plane waves in air, respectively below and beyond $f\idr{g}\edr{s}$ (also the asymptot of the other dispersion curves in air).
{\color[rgb]{0,0,1}\hdashrule[0.5ex]{5em}{1pt}{2.5mm 1mm}}~: $|k_{yz}\edr{a}(f)|$ for acoustical waves radiated by the main spatial component $k_x=k_p$ of the first propagating transverse mode in the wave-guides, starting at $f\idr{g}\edr{a}$ (see Eqs.~\ref{eq:fga} and \ref{eq:DispersAir}).\CR
{\color[rgb]{0,0,1}\hdashrule[0.5ex]{5em}{1pt}{.5mm 2mm}}: $|k_{yz}\edr{a}(f)|$ for acoustical waves radiated by the main spatial component $k_x=2\,k_p$ of the second propagating transverse mode in the wave-guides, starting at $2\,f\idr{g}\edr{a}$.%
}
\label{fig:dispers_guide}
\end{center}
\end{figure}

Waves in a plate radiate efficiently (that is: far from the plate plane) when their wavelength is larger than that in air (supersonic waves). For structural waves propagating in a direction with a dynamical rigidity $\overline{D}$, this occurs at a frequency \mbox{$f=\dfrac{c\idr{a}^2}{2\pi\overline{D}^{\,1/2}}$}. For an orthotropic plate, the transition from the subsonic to the supersonic regime is gradual, beginning at the frequency corresponding to the larger rigidity. For the values of the spruce parameters adopted above, the dynamical rigidities are larger in the  $x$-directions: \mbox{$\approx$ 150} in spruce plates (cut-off corners, usually ribless)  and \mbox{100 m$^4$s$^{-2}$} in the homogenised central zone. The coincidence frequencies are \mbox{1.5} and \mbox{1.8 kHz} respectively, both above the upper limit of the low-frequency regime. In this regime, the soundboard behaves therefore like a set of plates in their subsonic regime (analysed in~\cite{WAL1987}, for example). For the homogenised ribbed zone of the soundboard (main area of acoustical radiation), the lowest of the dispersion curves is represented in \Fig{dispers_guide} (thick solid red line), with $f_c$ defined as the lowest frequency corresponding to coincidence in this orthotropic plate:
\begin{equation}
f\idr{c}=\dfrac{c\idr{a}^2}{2\pi\overline{D_x\edr{H}}^{\,1/2}}
\end{equation}

Above $f\idr{g}\edr{s}$, the soundboard vibrates similarly to a set of three adjacent wave-guides. We suppose that the rest of the soundboard is at rest, more or less ensuring a baffle for the acoustical field. Acoustical radiation by ribbed panels has been studied extensively since the 60's and 70's by Heckl, Maidanik, Mead, Mace, and many others since. With regularly spaced ribs, the vibration extends all over the plane. The localised modal shapes are not known with precision in the $x$-direction but in any case, their spatial spectrum in this direction is maximum (with a more or less strong peak) at $k\edr{s}_x\approx\dfrac{m\pi}{p}$.

The structure-borne and the air-borne waves have the same spatial spectra in the $xy$-plane. Imposing a stationary field in the $x$-direction (with $k\edr{a}_x=m\pi/p$) yields the following dispersion equation for the acoustical planes waves radiated in a direction belonging to the $yz$-plane with wave number $k\edr{a}_{yz}$:
\begin{equation}
\left(k_{yz}\edr{a}\right)^2+\left(\dfrac{m\pi}{p}\right)^2=\dfrac{\omega^2}{c\idr{a}^2}
\label{eq:DispersAir}
\end{equation}

The first two dispersion branches (\mbox{$m=1$} and \mbox{$m=2$}) are drawn in \Fig{dispers_guide} (thin dashed and dotted blue curves). Defining
\begin{equation}
f\idr{g}\edr{a}\,=\,\dfrac{c\idr{a}}{2p}
\label{eq:fga}
\end{equation}
these dispersion branches intercept the $x$-axis at $f\idr{g}\edr{a}$ and $4\,f\idr{g}\edr{a}$. Their asymptots are the air dispersion line $k\edr{a}=2\pi\,f/c\idr{a}$ (thin dash-dot blue curve). Due to the  breadth of the $k_x\edr{s,a}$-spectrum (corresponding to localisation), these curves must be considered as the mean-lines of dispersion bands.

The $k_y$ components of the spatial spectra are also equal. The far-field acoustical radiation in the $z$-direction exists only if $|k\edr{a}_z|$ is positive. For waves propagating in the $y$-direction (which are radiated by the structural waves in the wave-guides), $|k\edr{a}_y|$ must be less than $|k\edr{a}_{yz}|$: this is the usual "supersonic condition" for the radiation of a structure-borne wave.

It is clear in \Fig{dispers_guide} that the subsonic or supersonic nature of the structural waves depends on the relative values of $f\idr{g}\edr{a}$ (given by \Eq{fga} and $f\idr{g}\edr{s}$ (given by \Eq{fgs}). We examine now what is the condition on $f$ for the structure-borne wave to be supersonic. With the following notations and normalisations (identical to \Eq{Normalisation} only for \mbox{$m=1$}):
\begin{equation}
\tilde{K}=\left(\dfrac{k}{k_p}\right)^2 \qquad\qquad\tilde{\Omega}=\left(\dfrac{\omega}{\omega\idr{g}\edr{s}}\right)^2
\end{equation}
the dispersion curves Eqs.~(\ref{eq:DispersWGnorm}) and (\ref{eq:DispersAir}), respectively in wave-guides and air, are transformed into:
\begin{align}
\tilde{K_y\edr{s}}\,&=\zeta\sqrt{\tilde{\Omega}\,-\,m^4(1-\gamma^2)}\,-\,m^2\,\zeta\,\gamma\label{eq:DispersWG12}\\
\tilde{K}\edr{a}_{yz}\,&=\,\dfrac{\tilde{\Omega}}{\tilde{\Omega}\idr{g}\edr{a}}\,-\,m^2\label{eq:DispersAirNorm}
\end{align}

\begin{figure}[ht!]
\includegraphics[width=\linewidth]{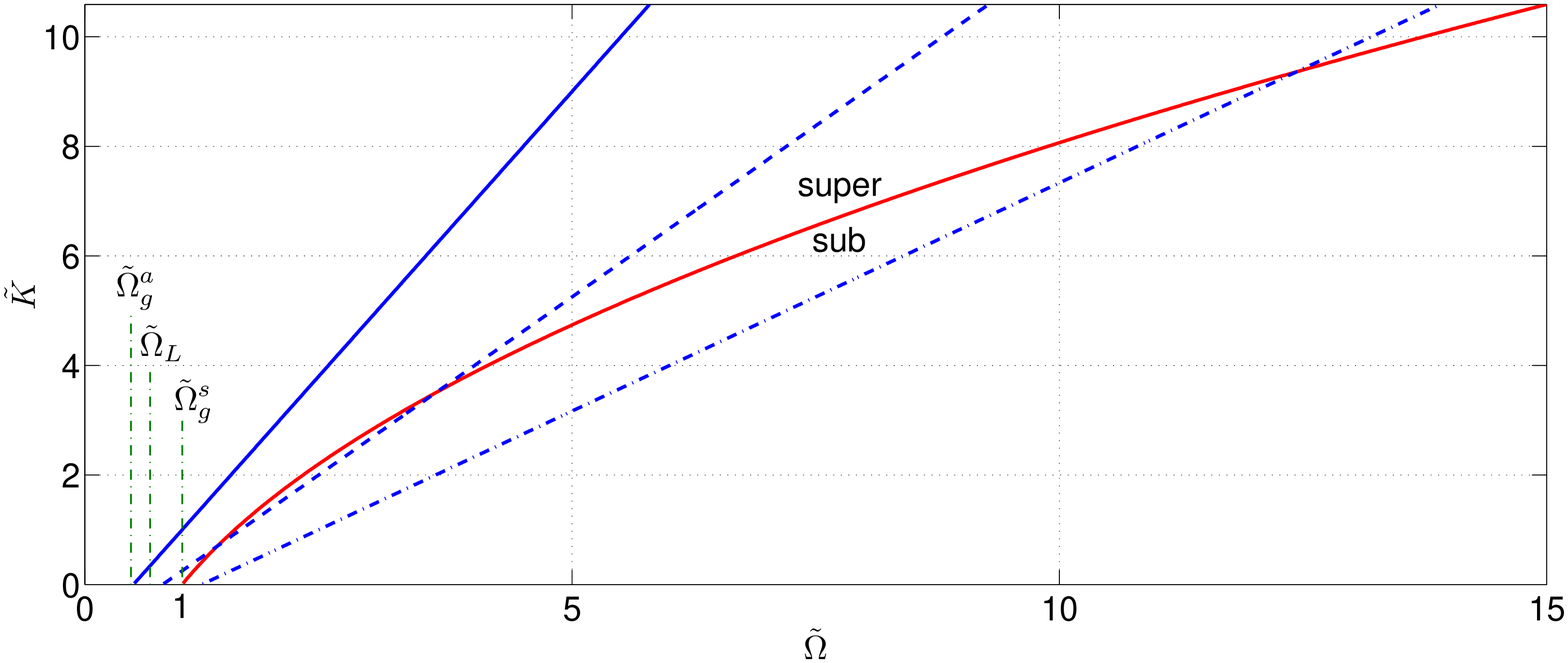}
\caption[aa]{Dispersion curves for structural and air-waves carried by the first transverse mode of the wave-guide ($m=1$), in normalised, squared coordinates $\tilde{\Omega}=\left(\omega/\omega\idr{g}\edr{s}\right)^2$ and $\tilde{K}=(k\,p/\pi)^2$.\CR
Red solid parabola: structural waves in a wave-guide (\Eq{DispersWG12}), with $x$-intercept at $\tilde{\Omega}=1$.
Blue straight lines: air waves (\Eq{DispersAirNorm}) with $x$-intercept at $\tilde{\Omega\idr{g}\edr{a}}$.\CR
The structural parameter $p/w$ determines the position of the blue line. $\tilde{\Omega\idr{L}}$ is the value of $\tilde{\Omega\idr{g}\edr{a}}$ for which the blue line is tangent to the red parabola. The acoustical radiation is efficient (supersonic waves) for frequencies at which a blue straight line is above the red parabola.\CR
Solid: $f\idr{g}\edr{a}<f\idr{L}$. Dashed: $f\idr{L}<f\idr{g}\edr{a}<f\idr{g}\edr{s}$. Dash-dot: $f\idr{g}\edr{a}>f\idr{g}\edr{s}$ (as usually observed in pianos).\CR
}
\label{fig:DispersSquare}
\end{figure}

The two equations are graphically represented in \Fig{DispersSquare} for \mbox{$m=1$} and various instances of $\tilde{\Omega}\idr{g}\edr{a}=\left(\dfrac{\omega\idr{g}\edr{a}}{\omega\idr{g}\edr{s}}\right)^2\,\propto\,\left(\dfrac{p}{w}\right)^2$. The red parabola represents \Eq{DispersWG12} and has a fixed $x$-intercept at $\tilde{\Omega}=1$. The straight blue lines represent \Eq{DispersAirNorm} and their $x$-intercept at $\tilde{\Omega}\idr{g}\edr{a}$ varies with the structural parameter $p/w$. Two cases must be distinguished.

(a) $\tilde{\Omega}\idr{g}\edr{a}>1\ \Leftrightarrow \ f\idr{g}\edr{a}>f\idr{g}\edr{s}$: dash-dotted line in \Fig{DispersSquare}, with only one intersection point between the dispersion curves. This is obtained with an inter-rib spacing of \mbox{$p=$ 13 cm} with a spruce panel of \mbox{$w\approx$ 8 mm} thick: $f\idr{g}\edr{a}$ is about 1.3~kHz, slightly above the frequency limit $f\idr{lim}\approx f\idr{g}\edr{s}$ of 1.1~kHz (see \Eq{fgs}), as shown in \Fig{dispers_guide} (dashed blue curve). The structural waves are subsonic below the frequency corresponding to this intersection point ($\tilde{\Omega}\approx 13$ in \Fig{DispersSquare}, $f\approx 7$~kHz in \Fig{dispers_guide}), and supersonic above\footnote{One notes here that the structural waves corresponding to $m=2$ are supersonic as soon as they appear.}. In other words, the subsonic regime occurring naturally in the low-frequency regime and up to $f\idr{g}\edr{a}$ in the high-frequency regime (evanescent waves: see \Eq{DispersAirNorm}) has been considerably extended by the ribbing of the soundboard. With a homogeneous plate equivalent to the ribbed soundboard (in plywood, for example), the supersonic regime would have ceased at \mbox{$f\idr{c}\approx$ 1.8 kHz}. Knowing how piano manufacturing has evolved, it is hard to think that this is the result of chance.

(b) $\tilde{\Omega}\idr{g}\edr{a}<1\ \Leftrightarrow \ f\idr{g}\edr{a}<f\idr{g}\edr{s}$: dashed and solid lines in \Fig{DispersSquare}. It can be seen graphically and demonstrated by some algebraic manipulations that there is one and only one value $\tilde{\Omega\idr{L}}\,\in\,[0,1]$ such that the air dispersion line is tangent to the wave-guide dispersion parabola. This case could be observed for low values of the structural parameter $p/w$. The two dispersion curves have:
\begin{itemize}
\item no intersection for $\tilde{\Omega\idr{g}\edr{a}}<\tilde{\Omega\idr{L}}$ (or \mbox{$f\idr{g}\edr{a}<f\idr{L}$}): solid blue line in \Fig{DispersSquare}. Assuming that $f\idr{c}$ remains above $f\idr{lim}$, structural waves would be  subsonic in the low-frequency regime and supersonic in the high-frequency regime.
\item two intersections for $\tilde{\Omega\idr{L}}<\tilde{\Omega\idr{g}\edr{a}}<1$ (or $f\idr{L}<f\idr{s}\edr{a}<f\idr{g}\edr{s}$): dashed line in \Fig{DispersSquare}. In this intermediate case, the radiation is alternatively subsonic and supersonic.
\end{itemize}

\subsection{Discussion}
The above observations may answer a puzzling question on piano manufacturing: why soundboards are not made out of plywood? Properly designed plywood would behave similarly in the low-frequency domain (homogeneous equivalent plate, with a low orthotropic ratio) and would radiate more efficiently above $f\idr{c}$, which might be considered as desirable on a musical instrument. A possible answer is that such a strong change in the radiation regime is undesirable since it would alter the homogeneity from note to note, both in level and in sustain and also introduce a sharp variation in the spectrum of each note. As shown above, the wave-guide regime extends the subsonic regime and may be a cure for this problem.


Dimensioning a piano model is done for specified nominal wood characteristics. If dimensioning is such that $f\idr{c}$ is adjusted too close to $f\idr{g}\edr{a}$, there is a risk that the dispersion in wood characteristics confers the undesirable radiating feature described above (case (b)) to some pianos of a manufacturing series. It is tempting to establish a connection between the frequency range where the radiation pattern of the piano may be strongly modified by the wave-guide phenomenon if it is not adjusted properly, and the so-called killer octave some manufacturers complain about. The transition between the two vibratory regimes of the soundboard and the induced non-uniformity of the acoustical radiation may explain why the sustain is so difficult to obtain for piano manufacturers around the fifth to sixth octave\footnote{See for example comments of the Fandrich Piano Company's piano maker~\cite{FAN1995}} ($\mathbf{C_6}\approx 1050 $~Hz).

Many attempts have been done to improve (or, more humbly, modify) the sound of the piano instrument by changing some features in the construction of its soundboard. However, very few have been carried along with acoustical measurements, and even less have been documented. The experimental study carried out by Conklin on a concert grand investigates the influence of ribbing on the sound radiation~\cite{CON1975}. Conklin built a soundboard with $39$ ribs (more than twice the usual number), reducing the spacing $p$ to a value of about 5-6~cm. The height of the ribs was the same as those of a normally-designed soundboard. Their width was changed to around $1.1$~cm, approximately half of the usual value, in order to keep almost the same rigidity and mass as that of a conventionally designed soundboard. It follows that $f\idr{g}\edr{s}$ reaches the highest frequency at which Conklin was interested, corresponding to the fundamental of the highest string of the piano: $\mathbf{C_8}\approx4200$~Hz. In his own words, Conklin's new soundboard "has improved uniformity of frequency response, improved and extended high frequency response, higher efficiency at higher frequencies, and improved tone quality". Of course, these conclusions need to be taken with some caution since no measurement was published and the soundboard was not available for third-parties' comments. With these modifications, the coincidence phenomenon is changed so that the supersonic radiation regime appears between approximately 1 and 4~kHz: favourable to the "high frequency response" and the "efficiency at higher frequencies" but unfavourable to sustain \ldots on which Conklin does not comment.

Suzuki~\cite{SUZ1986} measured the radiation efficiency of a baby-grand piano from 10~Hz to 5.4~kHz. It is interesting to note that there is no sensitive increase of the average radiation efficiency. A very smooth change appears in a so-called "transition range" of 1-1.6~kHz, decreasing slowly beyond. This absence of the sharp transition in radiation efficiency between the subsonic and the supersonic regimes is consistent with what has been found above.

%% file: CorrectionModalDensityOrthoV44.tex
\section{\appendixname: Modal density of a homogeneous plate -- Cases of non-special orthotropy and of arbitrary contour geometry}
\label{sec:AppModalOrtho}
For a plate of finite area $A$, the asymptotic modal density (reciprocal of the average difference between two consecutive resonance frequencies) of transverse waves is independent of the boundary conditions and, similarly, of the shape of the plate (see \cite{COU1953} for example\footnote{In his textbook \emph{Acoustics: an introduction to its physical principles and applications}, p.~293, Allan Pierce dates this result back to Weyl~\cite{WEY1912}.}). In low frequency, boundary conditions must be taken into account.

For a rectangular plate with an isotropic material of dynamical rigidity $\overline{D\idr{iso}}$, the asymptotic value and the boundary conditions correction for a rectangular plate with perimeter $L$ are~\cite{XIE2004}:
\begin{align}
n_{\infty,\text{iso}}&=\dfrac{A}{2\,\overline{D\idr{iso}}^{\,1/2}}\label{eq:ModalDensAsIso}\\
n\idr{iso}(f) &= n_{\infty,\text{iso}}\left(1+\epsilon\ \dfrac{L}{A}\dfrac{\overline{D\idr{iso}}^{\,1/4}}{\sqrt{2\pi\,f}}\right)\,= n_{\infty,\text{iso}}\left(1+\epsilon\ \dfrac{L}{\sqrt{4\pi A\overline{f}}}\right)\label{eq:OnePlateModalDensity}
\end{align}
where $\epsilon$ depends on the boundary conditions: -1/2 for the hinged case, -1 for the clamped case, +1 for the free case. The last expression makes use of the normalised frequency $\overline{f}=f\,n_{\infty,\text{iso}}$.

\begin{figure}[ht!]
\begin{center}
\includegraphics[width=0.35\linewidth]{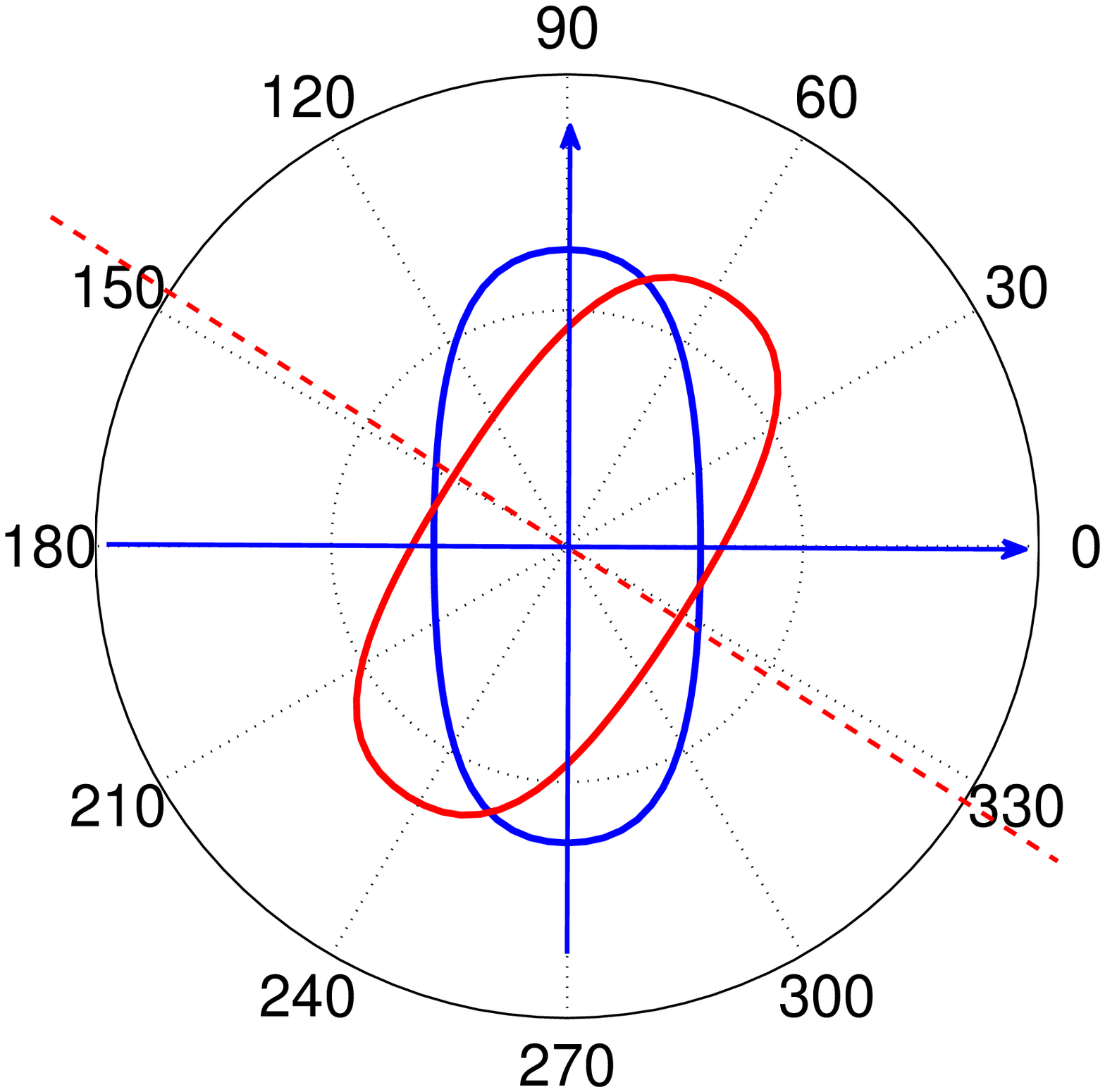}
\hspace{0.22\linewidth}
\includegraphics[width=0.35\linewidth]{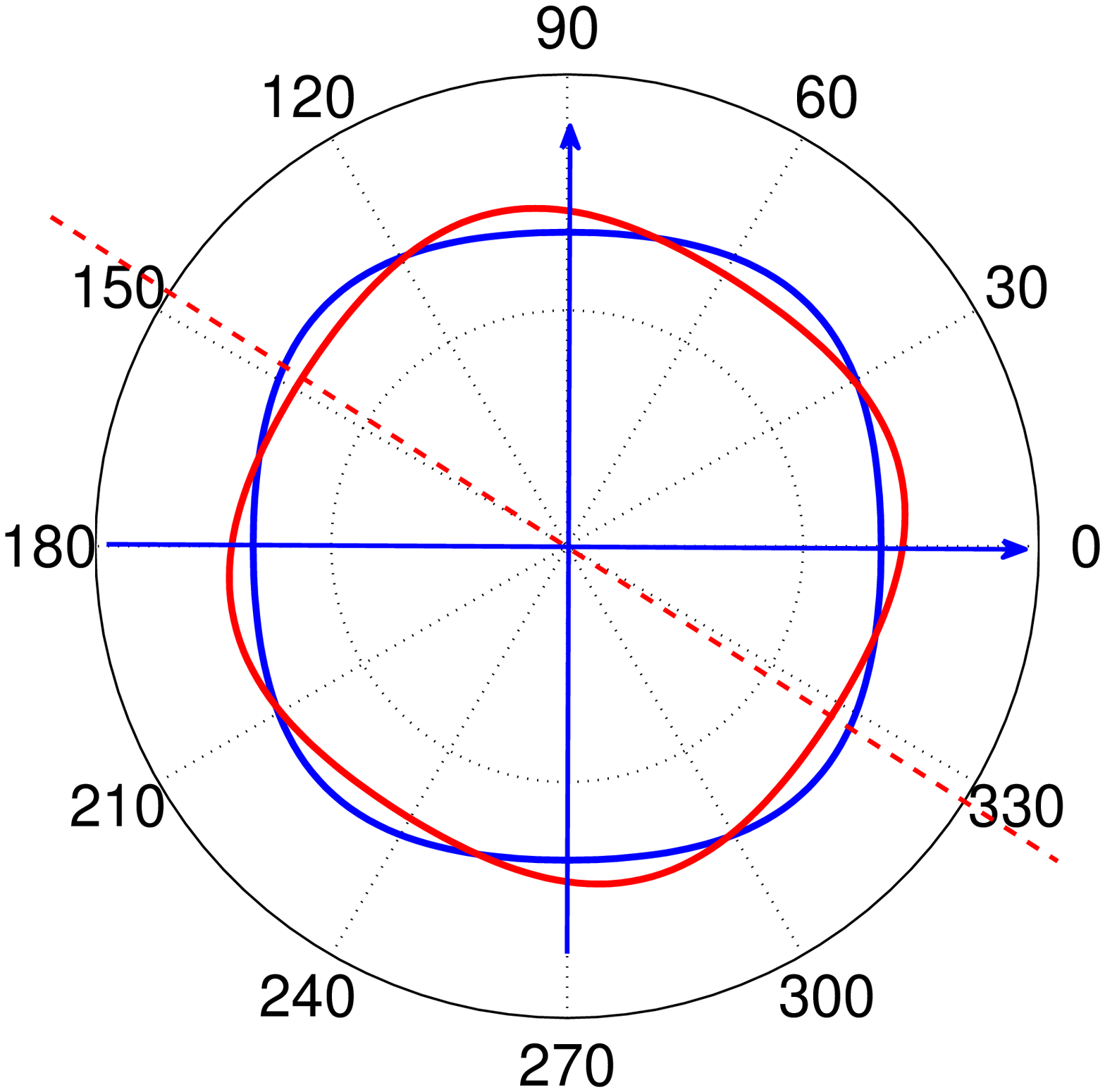}
\end{center}

\hspace{.2\linewidth}(a)\hspace{.55\linewidth}(b)
\caption[asd]{Constant frequency contours of a bidimensional orthotropic system (Sitka spruce plate).$\quad$ (a): in the wave-number $k$ plane, according to \Eq{DispersOrtho}; the blue axes correspond to $k_x$ and $k_y$.$\quad$ (b): in the modified $\kappa$ plane, according to \Eq{disperskappa}; the blue axes correspond to $\kappa|_{\theta=0}=D_1^{1/4}\,k_x$ and $\kappa|_{\theta=\pi/2}=D_3^{1/4}\,k_y$.\CR
{\color[rgb]{0,0,1}\hdashrule[.5ex]{3em}{1pt}{}}~: special orthotropy ($\theta_\perp=0$).$\quad$ {\color[rgb]{1,0,0}\hdashrule[.5ex]{3em}{1pt}{}}~: non-special orthotropy with $\theta_\perp\approx-32.5^\text{o}$ (corresponding to the case of our piano).%
}
\label{fig:angle_ortho}
\end{figure}

Although the literature provides all the necessary ingredients to obtain the asymptotic modal density of orthotropic rectangular plates with any orthotropic angle $\theta_\perp\neq0$ (defined as the angle between the long side of the rectangular plate and the main axis of orthotropy, see Fig.~\ref{fig:angle_ortho}), it seems to have been explicitly given in the case of special orthotropy only ($\theta_\perp=0$: plate sides $L_X$ and $L_Y$ parallel to the $x$- and $y$-directions respectively). The same observation applies to the boundary conditions correction at low-frequencies. This appendix aims at filling these small gaps. We also propose, without proof, a generalisation of the low-frequency correction for arbitrary geometry of the contour.

\subsection{Asymptotic modal density}
This section extends Wilkinson's work~\cite{WIL1968} done in the case of special orthotropy. %
%
With new polar coordinates $\kappa$ and $\psi$:
\begin{equation}
\left\{\begin{array}{ll}
\kappa\,\cos\psi&=k\,D_x^{1/4}\cos(\theta-\theta_\bot)\\
\kappa\,\sin\psi&=k\,D_y^{1/4}\sin(\theta-\theta_\bot)
\end{array}\right.
\end{equation}
the dispersion law \Eq{DispersOrtho} becomes:
\begin{equation}
\label{eq:disperskappa}
\kappa^4\,\left(\cos^4\psi+\cfrac{2D_{xy}}{\sqrt{D_xD_y}}\,\cos^2\psi\,\sin^2\psi+\sin^4\psi\right)=\rho\,h\,\omega^2
\end{equation}
and, as in~\cite{WIL1968}, can be factorised in:
\begin{equation}\label{eq:dispers_ortho_modif_angle}
\kappa^4(\omega,\psi)=\cfrac{\rho\,h\,\omega^2}{1-\alpha^2\sin^2(2\psi)}
\end{equation} 
where $\alpha^2$ is given in~	\eqref{eq:AlphaDef}. Adopting Courant~\cite{COU1953} and Bolotin~\cite{BOL1965} approaches, the asymptotic number of resonant modes $N(\omega)$ below a certain angular frequency $\omega$ is:
\begin{align}
N(\omega)&=\cfrac{L_X\,L_Y}{\pi^2(D_xD_y)^\frac{1}{4}}\int_0^{\kappa}\int_0^{\pi/2}{\kappa(\omega,\psi)\,\text{d}\psi\,\text{d}\kappa}=\cfrac{L_X\,L_Y}{2\pi^2(D_xD_y)^\frac{1}{4}}\int_0^{\pi/2}{\kappa^2(\omega,\psi)\,\text{d}\psi} \nonumber\\
\Rightarrow \quad
N(\omega)&=\cfrac{A\:\omega}{2\pi^2}\,\sqrt{\cfrac{\rho\,h}{D_x}}\,\left(\cfrac{D_x}{D_y}\right)^{\frac1{4}}\,F\left(\alpha\right)
\label{eq:nb_mode_ortho}\\
\text{with}\quad F(\alpha)&=\int_0^{\pi/2}{\left(1-\alpha^2\sin^2(2\psi)\right)^{-1/2}\text{d}\psi}
=\int_0^{\pi/2}{\left({1-\alpha^2\sin^2\psi}\right)^{-1/2}\text{d}\psi}
\nonumber
\end{align}

The last form of F is obtained using the even-parity and the $\pi$-periodicity of the function $\sin^2\psi$. 

By derivation of $N(\omega)$, the modal density is:
\begin{equation}
n_\infty(f)=\cfrac{\text{d}N(\omega)}{\text{d}\omega}\,\cfrac{\text{d}\omega}{\text{d}f}=\cfrac{A}{\pi}\,\sqrt{\cfrac{\rho\,h}{D_x}}\,\left(\cfrac{D_x}{D_y}\right)^{\frac1{4}}\,F\left(\alpha\right)=\cfrac{A}{2(\overline{D_x}\,\overline{D_y})^{1/4}}\,\cfrac{2F(\alpha)}{\pi}
\label{eq:AsModDensOrtho}
\end{equation}
where the last form is similar to \Eq{ModalDensAsIso}.
 
Finally, the result in the case of the non-special orthotropy is the same as that given by Wilkinson~\cite{WIL1968}. Since the angle of orthotropy $\theta_\bot$ is, in effect, the orientation of the plate boundaries with respect to the main axis of orthotropy, it was to be expected that $n_\infty$, being independent from boundary conditions, does not depend on $\theta_\bot$ either. For the general non-isotropic case, see Langley~\cite{LAN1996}.

\subsection{Low-frequency correction due to the boundary conditions}
We derive the correction to be brought in low-frequency to the modal density of the non-special orthotropic plate, due to the boundary conditions. The approach is based on the work of Xie~\emph{et~al.}~\cite{XIE2004} which yields Eq.~\eqref{eq:OnePlateModalDensity} in the case of an isotropic plate.

Introducing
\begin{equation}
\tilde{D}(\theta,\theta_\bot)=D_x\,\cos^4(\theta-\theta_\bot)+2\,D_{xy}\,\cos^2(\theta-\theta_\bot),\sin^2(\theta-\theta_\bot)+D_y\,\sin^4(\theta-\theta_\bot)
\end{equation}
transforms the dispersion law~\eqref{eq:DispersOrtho} into:
\begin{equation}
\tilde{D}(\theta,\theta_\bot)k^4(\omega,\theta)=\rho h\omega^2
\end{equation}

The correction terms on the mode count can be derived similarly to the isotropic case (exposed in details in \cite{XIE2004}) by adding or removing (depending of the type of boundary condition) the areas of two strips of modes along the main axes of the wavenumber diagram. Assuming that for simply supported boundary conditions, the wavenumbers may be approached by $m\dfrac{\pi}{L_X}$ and $n\dfrac{\pi}{L_Y}$ ($m$ and $n\in\mathbb N^*$), we obtain the number of modes and the modal density of a non-special orthotropic plate as follows:
\begin{align}
N(\omega)&=N_\infty({\omega})\,-\,\dfrac{k(\omega,0)\dfrac{\pi}{2L_Y}+k(\omega,\pi/2)\dfrac{\pi}{2L_X}\,+\,\dfrac{\pi}{2L_X}\dfrac{\pi}{2L_Y}}{\dfrac{\pi}{L_X}\dfrac{\pi}{L_Y}}\nonumber\\
&=N_\infty({\omega})\,-\,\dfrac{\left(\dfrac{\rho\,h}{\tilde{D}(0,\theta_\bot)}\right)^{1/4}\sqrt{\omega}\dfrac{\pi}{2L_Y}+\left(\dfrac{\rho\,h}{\tilde{D}(\pi/2,\theta_\bot)}\right)^{1/4}\sqrt{\omega}\dfrac{\pi}{2L_X}\,+\,\dfrac{\pi}{2L_X}\dfrac{\pi}{2L_Y}}{\dfrac{\pi}{L_X}\dfrac{\pi}{L_Y}}\nonumber\\
&=N_\infty({\omega})\,-\dfrac1{4\pi}(\rho\,h)^\frac1{4}\,\left[\dfrac{2L_X}{\tilde{D}(0,\theta_\bot)^{1/4}}+\dfrac{2L_Y}{\tilde{D}(\pi/2,\theta_\bot)^{1/4}}\right]\,\sqrt{\omega}\\
n(f)&=n_\infty\,-\dfrac1{\sqrt{32\pi}}(\rho\,h)^{1/4}\,\left[\dfrac{2L_X}{\tilde{D}(0,\theta_\bot)^{1/4}}+\dfrac{2L_Y}{\tilde{D}(\pi/2,\theta_\bot)^{1/4}}\right]\,f^{-1/2} \label{eq:northoangle_s} 
\end{align}

With wavenumbers approximated by $\left(m+\dfrac1{2}\right)\dfrac{\pi}{L_X}$ and $\left(n+\dfrac1{2}\right)\dfrac{\pi}{L_Y}$, we obtain in the case of the clamped non-special orthotropic plate:
\begin{equation}	n(f)=n_\infty\,-\cfrac1{\sqrt{8\pi}}(\rho\,h)^{1/4}\,\left[\dfrac{2L_X}{\tilde{D}(0,\theta_\bot)^{1/4}}+\dfrac{2L_Y}{\tilde{D}(\pi/2,\theta_\bot)^{1/4}}\right]\,f^{-1/2}
\label{eq:northoangle_e}
\end{equation}

In the case of free boundary conditions, with wavenumbers approximated by $\dfrac{m\pi}{2L_X}$ and $\dfrac{n\pi}{2L_Y}$, and accounting for the rigid and beam-modes~\cite{XIE2004} yields:
\begin{equation}
n(f)=n_\infty\,+\cfrac1{\sqrt{8\pi}}(\rho\,h)^{1/4}\,\left[\dfrac{2L_X}{\tilde{D}(0,\theta_\bot)^{1/4}}+\dfrac{2L_Y}{\tilde{D}(\pi/2,\theta_\bot)^{1/4}}\right]\,f^{-1/2}
\label{eq:northoangle_l} 
\end{equation}

These formula can be written in more compact forms, similar to \Eq{OnePlateModalDensity}:
\begin{align}
n(f)&= n_\infty\left[1+\cfrac{\epsilon}{\sqrt{2\pi f}}\,\cfrac{(D_xD_y)^{1/4}}{A}\,\cfrac{\pi}{2F(\alpha)}\,\left(\cfrac{2L_X}{\tilde{D}(0,\theta_\bot)^{1/4}}+\cfrac{2L_Y}{\tilde{D}(\pi/2,\theta_\bot)^{1/4}}\right)\right]\\
\Rightarrow \quad n(f)&=n_\infty\left(1+\cfrac{\epsilon\tilde{L}}{\sqrt{4\pi A\overline{f}}}\right)\label{eq:ModalDensityGen}
\end{align}
with $n_\infty$ given by \Eq{AsModDensOrtho}, $\overline{f}=f\,n_\infty$, $\epsilon=$ given as in \Eq{OnePlateModalDensity} and 
\begin{equation}
\tilde{L}=\sqrt{\cfrac{\pi}{2F(\alpha)}}(D_xD_y)^{1/8}\left(\cfrac{2L_X}{\tilde{D}(0,\theta_\bot)^{1/4}}+\cfrac{2L_Y}{\tilde{D}(\pi/2,\theta_\bot)^{1/4}}\right)
\label{eq:ContourCorrRect}
\end{equation}

Naturally, Eqs.~(\ref{eq:northoangle_s})--(\ref{eq:northoangle_l}) yield Eq.~(\ref{eq:OnePlateModalDensity}) when $D=D_x=D_y=D_{xy}$ (isotropic plate), for any $\theta_\bot$. One notes also that for special orthotropy the expression in parentheses in \Eq{ContourCorrRect} becomes:  $\cfrac{2L_X}{{D_x}^{1/4}}+\cfrac{2L_Y}{{D_y}^{1/4}}$ when $\theta_\perp=a\pi$ (with $a\in\mathbb{N}$) and becomes $\cfrac{2L_X}{{D_y}^{1/4}}+\cfrac{2L_Y}{{D_x}^{1/4}}$ for inversed axis of orthotropy (that is when $\theta_\perp=\pi/2+a\pi$).

We form the hypothesis that \Eq{ContourCorrRect} can be generalised to any shape of $A$:
\begin{align}
\tilde{L}&=\sqrt{\cfrac{\pi}{2F(\alpha)}}\mathop{\mathlarger{\mathlarger{\oint}}}_L{\cfrac{(D_xD_y)^{1/8}}{\tilde{D}(\Theta)^{1/4}}\dd s}\\
&=\sqrt{\cfrac{\pi}{2F(\alpha)}}\mathop{\mathlarger{\mathlarger{\oint}}}_L{\left(\cfrac{\zeta}{\zeta^2\cos^4\Theta\,+\,2\zeta\cos^2\Theta\sin^2\Theta\,+\,\sin^4\Theta}\right)^{1/4}\dd s}\label{eq:ContourCorrGen}
\end{align}
where $\Theta=\theta-\theta_\bot$ is the polar angle of $\dd s$ and $\zeta^2=D_x/D_y$ (see \Eq{zetaDef}). In the case of isotropy, $\tilde{L}$ would simply be the perimeter of $A$.

%% file: Equiv_dynrigid.tex
\section{\appendixname: Equivalent isotropic dynamical rigidity of piano soundboards}
\label{sec:equiv_dynrigid}

It has been suggested in the literature that the ribs and the bridges compensate globally the anisotropy of spruce~\cite{LIE1979,CON1996_2}. This hypothesis is tested here by computing a dynamical rigidity $\overline{D\idr{iso}}$ for different piano soundboards. The literature on piano soundboard offers data (reported in~\cite{EGE2009_2bib}, p.~17, 18) that can be analysed according to a simple equivalent isotropic plate model: Suzuki~\cite{SUZ1986}, Dérogis~\cite{DER1997}, Berthaut \emph{et al.}~\cite{BER2003}, and ourselves~\cite{Ege2013a} 
have published modal frequencies (below 500~Hz or less for the three first authors). It is out of question to apply to these instruments the full sub-plate model developed in Section~\ref{sec:LowFreq} since only the overall dimensions of the soundboards are reported.

We propose here to model each soundboard as a homogeneous isotropic rectangular plate with the same area $A$, modal density $n\idr{iso}(f)$ and clamped boundary conditions along the perimeter $L$. Combining Eqs.~\ref{eq:ModalDensAsIso}--\ref{eq:OnePlateModalDensity} (with $\epsilon=-1$) one notes that the quantity $r=\overline{D\idr{iso}}^{\,1/4}$ obeys the following equation:
\begin{equation}
2\,n\idr{iso}(f)\,r^2\,+\,\dfrac{L}{\sqrt{2\pi\,f}}\,r\,-A\,=0
\label{eq:rEval}
\end{equation}

\begin{figure}
\begin{center}
\includegraphics[width=\linewidth]{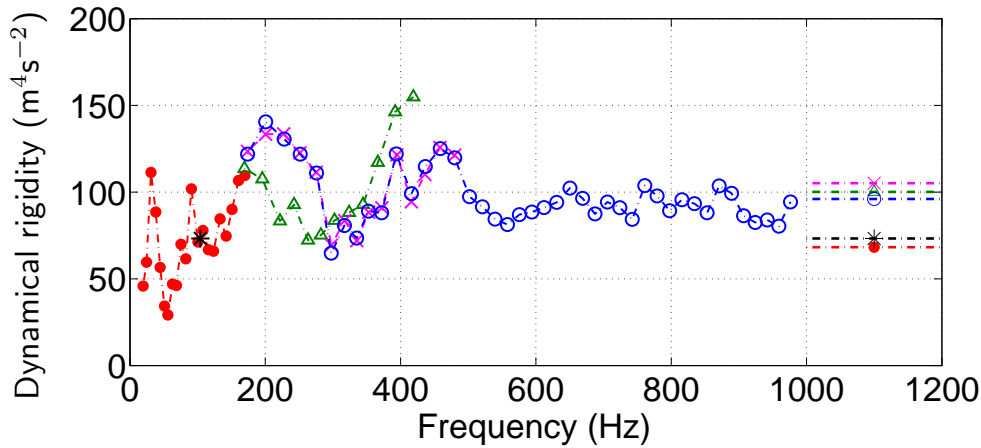}
\caption[aaa]{Estimations of the dynamical rigidities $\overline{D\idr{iso}}$ of the homogeneous isotropic plates equivalent to different piano soundboards explored in the literature. Estimations are based on the measurements of the modal densities as the reciprocal of the moving average on six successive intermodal-spacings. The horizontal lines at the right side of the figure correspond to the average value $<\overline{D\idr{iso}}>$ of each series of points.\CR
{\color[rgb]{1,0,1}$\times$}: Ege \emph{et al.}~\cite{Ege2013a}, impact excitation on an upright piano,  $<\overline{D\idr{iso}}>=105~\text{m}^{4}~\text{s}^{-2}$.
{\color[rgb]{0,0,1}$\circ$}: Ege \emph{et al.}~\cite{Ege2013a}, acoustical excitation (same piano, wider frequency range), $<\overline{D\idr{iso}}>=96~\text{m}^{4}~\text{s}^{-2}$.\CR
{\color[rgb]{0,0.5,0}\tiny{$\Delta$}}: Dérogis~\cite{DER1997}, upright piano, $<\overline{D\idr{iso}}>=100~\text{m}^{4}~\text{s}^{-2}$.\CR
{\color[rgb]{1,0,0}\tiny{$\bullet$}}: Berthaut~\emph{et~al.}~\cite{BER2003}, baby grand piano, $<\overline{D\idr{iso}}>=68~\text{m}^{4}~\text{s}^{-2}$.\CR
{\scriptsize{$\ast$}}: Suzuki~\cite{SUZ1986}, baby grand piano, one estimation only, due to the low number of reported modes, $\overline{D\idr{iso}}=73~\text{m}^{4}~\text{s}^{-2}$.%
}
\label{fig:Rigidity_Fit}
\end{center}
\end{figure}

The modal density of each soundboard has been estimated as the reciprocal of the moving average on six successive intermodal-spacings. The solutions of Eq.~\ref{eq:rEval} are reported in Fig.~\ref{fig:Rigidity_Fit}. It appears that the dynamical rigidities of the two uprights (Dérogis' and ours) and the two baby-grands (Suzuki's and Berthaut's) are very similar.

What would possibly constitute a manufacturing rule may be considered as what piano makers have come to achieve with the dimensioning of the many parts of a soundboard, over decades of empiricism or, possibly, what they kept from making habits on earlier keyboard instruments.

\renewcommand{\figureplace}{%
\begin{center}
[\figurename~\thepostfig\ about here (with ref.~\cite{Ege2013a,DER1997,BER2003,SUZ1986}).]
\end{center}}

\renewcommand{\figureplace}{%
\begin{center}
[\figurename~\thepostfig\ about here.]
\end{center}}